\documentclass[11pt]{article}%

\usepackage{amsmath,amssymb,amsfonts}
\usepackage{calc}
\usepackage{dsfont}
\usepackage{slashed}
\usepackage{cite}
\usepackage{epsfig}
\usepackage{slashed}
\usepackage{multirow}
\usepackage{color}
\usepackage[bottom]{footmisc}
\usepackage{ulem}

\newcommand{\beq}{\begin{eqnarray}}
\newcommand{\eeq}{\end{eqnarray}}

\newcommand{\s}{\newline \vspace*{-3.5mm}}

\makeatletter
\def\section{\@startsection {section}{1}{\z@}{-3.5ex plus -1ex minus 
-.2ex}{2.3ex plus .2ex}{\large\bf}}
\def\subsection{\@startsection{subsection}{2}{\z@}{-3.25ex plus -1ex
minus -.2ex}{1.5ex plus .2ex}{\normalsize\bf}}
\makeatother
\newcommand{\captionfonts}{\small}
\makeatletter  
\long\def\@makecaption#1#2{%
  \vskip\abovecaptionskip
  \sbox\@tempboxa{{\captionfonts #1: #2}}%
  \ifdim \wd\@tempboxa >\hsize
    {\captionfonts #1: #2\par}
  \else
    \hbox to\hsize{\hfil\box\@tempboxa\hfil}%
  \fi
  \vskip\belowcaptionskip}
\makeatother   

\renewcommand{\d}{{\rm d}}
\renewcommand{\bar}{\overline}

\newcommand{\sssty}[1]{\scriptscriptstyle#1}
\newcommand{\f}[2]{\frac{#1}{#2}}
\begin{document}

\title{
\vspace*{-3cm}
\phantom{h} \hfill\mbox{\small KA-TP-05-2016}\\[-1.1cm]
\phantom{h} \hfill\mbox{\small PSI-PR-16-02}\\[-1.1cm]
\phantom{h} \hfill\mbox{\small RM3-TH/16-3}
\\[1cm]
\textbf{Signs of Composite Higgs Pair Production} \\
\textbf{at Next-to-Leading Order}}

\date{}
\author{
{\bf R.~Gr\"ober$^{\,1}$}\footnote{E-mail: \texttt{groeber@roma3.infn.it}}, 
{\bf M.~M\"uhlleitner$^{\,2}$}\footnote{E-mail:
  \texttt{milada.muehlleitner@kit.edu}} $\;$and {\bf
  M.~Spira$^{\,3}$}\footnote{E-mail: \texttt{michael.spira@psi.ch}} 
\\[9mm]
{\small\it
$^1$ INFN, Sezione di Roma Tre, Via della Vasca Navale 84, I-00146
Roma, Italy}\\[3mm] 
{\small\it
$^2$Institute for Theoretical Physics, Karlsruhe Institute of Technology,} \\
{\small\it 76128 Karlsruhe, Germany}\\[3mm]
{\small\it
$^3$ Paul Scherrer Institute, CH-5323 Villigen PSI, Switzerland}}

\maketitle

\begin{abstract}
\noindent
In composite Higgs models the Higgs boson arises as a pseudo-Goldstone
boson from a strongly-interacting sector. Fermion mass generation is
possible through partial compositeness accompanied by the appearance
of new heavy fermionic resonances. The Higgs couplings to the
Standard Model (SM) particles and between the Higgs bosons 
themselves are modified with respect to the SM. Higgs pair production is
sensitive to the trilinear Higgs self-coupling but also to anomalous
couplings like the novel 2-Higgs-2-fermion coupling emerging in
composite Higgs models. The QCD corrections to SM Higgs boson pair
production are known to be large. In this paper we compute, in the
limit of heavy loop particle masses, the next-to-leading order (NLO) QCD
corrections to Higgs pair production in composite Higgs models without
and with new heavy fermions. The relative QCD corrections are found to be almost
insensitive both to the compositeness of the Higgs boson and to the details of
the heavy fermion spectrum, since the leading order cross section dominantly
factorizes. With the obtained results we investigate the question if,
taking into account Higgs coupling constraints, new physics could
first be seen in Higgs pair production. We find this to be the case in
the high-luminosity option of the LHC for composite Higgs models with
heavy fermions. We also investigate the invariant mass
distributions at NLO QCD. While they are sensitive to the Higgs
non-linearities and hence anomalous couplings, the influence of the
heavy fermions is much less pronounced.
\end{abstract}
\thispagestyle{empty}
\vfill
\newpage
\setcounter{page}{1}

\section{Introduction}
The LHC Higgs data of Run 1 suggest that the scalar particle observed
by the LHC experiments ATLAS and CMS in 2012 \cite{:2012gk,:2012gu} is
compatible with the Higgs boson of the Standard Model (SM). The
non-vanishing vacuum expectation value (VEV) $v$ of the $SU(2)$
Higgs doublet field $\phi$ in the ground state is crucial for the
mechanism of electroweak symmetry breaking (EWSB)
\cite{Higgs:1964pj}. It it is induced by the Higgs potential  
\beq
V = \lambda \bigg[ \phi^\dagger \phi - \frac{v^2}{2} \bigg]^2 \;.
\eeq
Introducing the Higgs field in the unitary gauge, $\phi = 
(0, [v+H]/\sqrt{2})^T$, it reads
\beq
V = \frac{M_H^2}{2} H^2 + \frac{\lambda_{HHH}}{3!} H^3 +
\frac{\lambda_{HHHH}}{4!} H^4 \;.
\eeq
In the SM the trilinear and quartic Higgs self-couplings
are uniquely determined in terms of the Higgs boson mass $M_H =
\sqrt{2\lambda} v$,
\beq
\lambda_{HHH} = \frac{3M_H^2}{v} \qquad \mbox{and} \qquad 
\lambda_{HHHH} = \frac{3 M_H^2}{v^2} \;, \label{eq:lammassrel}
\eeq
with $v \approx 246$~GeV. The experimental verification of the form of
the Higgs potential through the measurement of the Higgs
self-couplings is the final step in the program aimed to test the
mechanism of EWSB. The Higgs self-couplings
are accessible in multi-Higgs production processes
\cite{Dawson:1998py,Djouadi:1999gv,Djouadi:1999rca,thesis}. While
previous studies 
\cite{Baur:2003gpa,Baur:2002rb,Baur:2003gp,Dolan:2012rv,Papaefstathiou:2012qe,Baglio:2012np,Goertz:2013kp,Yao:2013ika,Barr:2013tda,Dolan:2013rja,Barger:2013jfa,deLima:2014dta,Englert:2014uqa,Wardrope:2014kya,Li:2015yia,Lu:2015jza,Dolan:2015zja,Cao:2015oxx,Behr:2015oqq} showed that the
probe of the trilinear Higgs self-coupling 
in Higgs pair production should be possible at the
high-luminosity LHC, although it is experimentally very
challenging, the quartic Higgs self-interaction is out of reach. The
cross section of triple Higgs production giving access to this
coupling suffers from too low signal rates fighting against a large
background \cite{Djouadi:1999gv,thesis,Plehn:2005nk}. The relations in
Eq.~(\ref{eq:lammassrel}) do not hold in models beyond the SM
(BSM), and this would manifest itself in the Higgs pair production
process. In general, however, new physics (NP) not only affects the 
value of the Higgs self-coupling, but also other couplings involved in the
Higgs pair production process.\footnote{Note, however, that in 
Ref.~\cite{Dermisek:2013pta} a model is discussed, where only the Higgs 
self-couplings are modified with respect to the SM via loop
corrections of an invisible new state.} 
An approach that allows to smoothly depart from
the SM in a consistent and model-independent way is offered by the
effective field theory 
(EFT) framework based on higher dimensional operators which are 
added to the SM Lagrangian with coefficients that are suppressed by
the typical scale $\Lambda$ where NP becomes
relevant
\cite{Burges:1983zg,Leung:1984ni,Buchmuller:1985jz,Hagiwara:1993ck,Grzadkowski:2010es}. 
These higher dimensional operators modify the couplings involved in
Higgs pair production, such as the trilinear Higgs
self-coupling and the Higgs Yukawa couplings. Additionally they
give rise to novel couplings, like a 2-Higgs-2-fermion coupling, that
can have a significant effect on the process.
While the trilinear Higgs 
self-coupling has not been delimited experimentally yet\footnote{In
  some NP models, the trilinear Higgs self-coupling can still deviate 
significantly from the SM expectations
\cite{Buttazzo:2015bka,Huang:2015tdv}.}, the Higgs couplings to the 
SM particles have been constrained by the LHC data and in particular the
Higgs couplings to the massive gauge bosons. An interesting question
to ask is, while taking into account the information on the Higgs properties
gathered at the LHC, if it could be that despite the Higgs boson behaving
SM-like, we see NP emerging in Higgs pair production? And if
so, could it even be, that we see NP before having any other direct
hints {\it e.g.}~from new resonances or indirect hints from {\it e.g.}~Higgs
coupling measurements? \s

Previous works have applied the EFT approach to investigate BSM
effects in Higgs pair production.\footnote{For effects of NP on the
  trilinear Higgs self-coupling and/or Higgs pair production within
  specific BSM models in recent studies, see
  {\it e.g.}~\cite{Grober:2010yv,Dawson:2012mk,Cao:2013si,Nhung:2013lpa,Ellwanger:2013ova,Han:2013sga,No:2013wsa,Arhrib:2013oia,Heng:2013cya,Barradas-Guevara:2014yoa,Efrati:2014uta,Baglio:2014nea,Hespel:2014sla,King:2014xwa,Barger:2014qva,Yang:2014gca,Liang-Wen:2014fla,Chen:2014ask,Liu:2015mza,Martin-Lozano:2015dja,Godunov:2015nea,Wu:2015nba,Dawson:2015haa,Batell:2015koa,Ginzburg:2015yva,Zhang:2015mnh,Costa:2015llh,Agostini:2016vze,Zhou:2016nkd}.}
A study of the effects of genuine dimension-six operators in Higgs pair
production can be found in Ref.~\cite{Barger:2003rs}. Anomalous
couplings in Higgs pair production have been investigated in
\cite{Contino:2012xk,Nishiwaki:2013cma,Contino:2013gna,Chen:2014xra}. 
In \cite{Goertz:2014qta,Azatov:2015oxa,He:2015spf} the EFT was applied
to investigate the prospects of probing the trilinear Higgs
self-coupling at the LHC.  
Reference \cite{Edelhaeuser:2015zra} on the other hand addressed the question on
the range of validity of the EFT approach for Higgs pair production by
using the universal extra dimension model. \s

The dominant Higgs pair production process at the LHC is gluon fusion,
$gg \to HH$, which is mediated by loops of heavy fermions. It can be
modified due to NP via deviations in the trilinear Higgs
self-coupling, in the Higgs to fermion couplings, via 
new couplings such as a direct coupling of two fermions to two Higgs
bosons, new particles like {\it e.g.}~heavy quark partners in the
loop, or additional (virtual) Higgs bosons, splitting into two lighter final
state Higgs bosons. 
The purpose of this paper is to address the question of whether it will be 
possible to see deviations from the SM for the first time in
non-resonant Higgs pair 
production processes by considering explicit models. 
It has been found that large deviations from SM Higgs pair
  production can arise in composite Higgs models, which is mainly due
  to the novel 2-Higgs-2-fermion coupling \cite{Grober:2010yv,Gillioz:2012se}. 
In this paper, we will hence focus on this class of models. 
We assume that no deviations with respect to the SM are seen in any of
the LHC Higgs coupling analyses, {\it i.e.}~that the deviations in 
the standard Higgs couplings due to NP are below the expected experimental
sensitivity, for the case of the LHC high-energy Run 2 and for the
high-luminosity option of the LHC. Additionally, we assume that no NP
will be observed in direct searches or indirect measurements. The
prospects of NP emerging from composite Higgs models for the first
time in non-resonant Higgs pair production from gluon fusion
are analyzed under these conditions. Our analysis is complementary to
previous works \cite{Gupta:2013zza,Efrati:2014uta}, which focused on deviations
in Higgs pair production due to modifications in the trilinear Higgs
coupling. In Ref.~\cite{Gupta:2013zza} the question is investigated on how
well the trilinear Higgs coupling needs to be measured in various
scenarios to be able to probe NP. The main focus
of Ref.~\cite{Efrati:2014uta} is on how to combine a deviation  
in the trilinear Higgs coupling with other Higgs coupling measurements to 
support certain BSM extensions. \s

Gluon fusion into Higgs pairs exhibits large QCD corrections. In 
Ref.~\cite{Dawson:1998py}, the next-to-leading order (NLO) QCD corrections were 
computed in the large top mass approximation and found to be 
of $\mathcal{O}(90\%)$ at $\sqrt{s}=14$~TeV for a Higgs boson
mass of 125 GeV. The effects of finite top quark masses have been analyzed in
\cite{Dawson:2012mk,Grigo:2013rya,Grigo:2013xya,Frederix:2014hta,Grigo:2014oqa,Maltoni:2014eza}. While
the $m_t \to \infty$ approximation exhibits uncertainties of order 20\% on
the leading order (LO) cross section at $\sqrt{s}= 14$~TeV for a
light Higgs boson 
\cite{Glover:1987nx,Plehn:1996wb,Gillioz:2012se} and badly fails 
to reproduce the differential distributions \cite{Baur:2002rb}, the
uncertainty on the $K$-factor, {\it i.e.}~the ratio between the
loop-corrected and the LO cross section, is much smaller due to the
fact that in the dominant soft and collinear contributions the full LO cross section can
be factored out. The 
next-to-next-to-leading order (NNLO) corrections have been provided 
by \cite{deFlorian:2013uza,deFlorian:2013jea,Grigo:2014jma} in the heavy top
mass limit. The finite top mass effects have been estimated to be of
about $10\%$ at NLO and $\sim 5$\% at NNLO \cite{Grigo:2015dia}. 
Soft gluon resummation at next-to-leading logarithmic order has been
performed in \cite{Shao:2013bz} and has been extended recently to the 
next-to-next-to-leading logarithmic level in
\cite{deFlorian:2015moa}. First results
towards a fully differential NLO calculation have been provided in
\cite{Maltoni:2014eza,Frederix:2014hta}. 
For a precise determination of the accessibility of BSM effects in gluon fusion 
to a Higgs pair, the NLO QCD corrections are essential and need to be computed 
in the context of these models. They have been provided in the
large loop particle mass limit for the singlet-extended SM
\cite{Dawson:2015haa}, for 
the 2-Higgs-doublet model \cite{Hespel:2014sla} and for the MSSM
\cite{Dawson:1998py,Agostini:2016vze}.\footnote{Reference
  \cite{Agostini:2016vze} also shows how the provided results can be
  adapted to the Next-to-Minimal Supersymmetric extension of the SM.} 
In the same limit, the NLO QCD corrections
including dimension-6 operators have been computed in
\cite{Grober:2015cwa}. In this work, we calculate for the first time the 
NLO QCD corrections in the large loop particle mass limit for models
with vector-like fermions such as composite Higgs models. \s

The paper is organized as follows. In Section~\ref{sec:chm} we
briefly introduce composite Higgs models. In section~\ref{sec:NLOQCD}
we present the NLO QCD corrections to the gluon fusion process in the
framework of composite Higgs models including vector-like
fermions. In the subsequent sections we analyze  
whether a possible deviation from the SM signal could be seen or not at the
LHC Run 2 with an integrated luminosity of $300\;\text{fb}^{-1}$
and/or the high-luminosity LHC with an integrated luminosity of
$3000\;\text{fb}^{-1}$ for different models: in
section~\ref{sec:numerical1} for the composite Higgs models MCHM4 and
MCHM5, and in section~\ref{sec:numerical2} for a composite Higgs model
with one multiplet of fermionic resonances below the cut-off. 
In section \ref{sec:invmasssdist} we discuss the invariant mass
distributions with and without the inclusion of the new fermions. We
conclude in section~\ref{sec:concl}.  

\section{Composite Higgs Models \label{sec:chm}}
In composite Higgs models the Higgs boson arises as a pseudo-Nambu Goldstone 
boson of a strongly interacting sector 
\cite{Kaplan:1983fs,Dimopoulos:1981xc,Banks:1984gj,Kaplan:1983sm,Georgi:1984ef,Georgi:1984af,Dugan:1984hq}. A
global symmetry is broken at the scale $f$ to a subgroup containing at
least the SM gauge group. The new strongly-interacting sector can
be characterized by a mass scale $m_\rho$ and a coupling
$g_\rho$, with $f=m_\rho/g_\rho$. An effective low-energy description of such
models is provided by the Strongly Interacting Light Higgs  
(SILH) Lagrangian~\cite{Giudice:2007fh}, which, in addition to the SM 
Lagrangian, contains higher dimensional operators including the SM
Higgs doublet $\phi$ to account for the composite nature of the Higgs
boson. Listing only the operators relevant for Higgs pair production
by gluon fusion, the SILH Lagrangian reads\footnote{We have not
  included the chromomagnetic 
  dipole moment operator which modifies the interactions between the
  gluons, the top quark and the Higgs boson and can be expected to be
  of moderate size \cite{Degrande:2012gr}.}  
 \beq
 \mathcal{L}^{\text{SILH}}&\supset& \f{c_{\sssty{H}}}{2 
f^2}\partial^{\mu}\left(\phi^{\dag}\phi\right)\partial_{\mu}\left(\phi^{\dag}
\phi\right)
-\f{c_6\lambda}{f^2}\left(\phi^{\dag}\phi\right)^3 \nonumber \\
&& 
+\left[\left(\f{c_u y_u}{f^2}\phi^{\dagger}\phi \,\bar{Q}_{\sssty{L}}\phi^c 
u_{\sssty{R}}+\f{c_d y_d}{f^2}\phi^{\dagger}\phi \,\bar{Q}_{\sssty{L}}\phi
d_{\sssty{R}} \right) +h.c.\right] \nonumber \\
&& +\f{c_{g} \alpha_{s}}{4\pi\,f^2}\f{y_t^2}{g_{\rho}^2}\phi^{\dag}\phi\, 
G^a_{\mu\nu}G^{a\mu\nu}\,,  \label{eq:silh}
\eeq
with the Yukawa couplings $y_q=\sqrt{2} m_q/v$
($q=u,d$), where $m_q$ 
denotes the quark mass, $\lambda$ the quartic Higgs coupling and
$\alpha_s = g_s^2/(4\pi)$ the strong coupling constant in
terms of the $SU(3)_c$ gauge coupling $g_s$.\footnote{The relation
    between the 
  coefficients $c$ and the coefficients $\bar{c}$ in Eq.~(2.1) of
  Ref.~\cite{Grober:2015cwa} is $\bar{c}_x = c_x \xi$ ($x=H,6$),
  $\bar{c}_u=c_u \xi$ and $\bar{c}_g = \alpha_2/(16\pi) y_t^2/g_\rho^2
  c_g \xi$ with $\xi=v^2/f^2$ and $\alpha_2 = \sqrt{2} G_F m_W^2/\pi$
  in terms of the Fermi constant $G_F$ and the $W$ boson mass
  $m_W$.} Here $Q_L$ denotes the left-handed quark doublet.
The effective Lagrangian  
accounts for several effects that can occur in Higgs pair production via gluon 
fusion in composite Higgs models: a shift in the trilinear Higgs self-coupling
and in the Higgs couplings to the fermions, a novel coupling of two
fermions to two Higgs bosons and additional new fermions in the
loops. The latter effect is encoded in the effective operator with the
gluon field strength tensors $G_{\mu\nu}$ coupling directly to the
Higgs doublet $\phi$. 
While the SILH Lagrangian Eq.~(\ref{eq:silh}) is a valid description
for small values of $\xi = (v/f)^2$, larger values require a
resummation of the series in $\xi$. This is provided by explicit models
built in five-dimensional warped space. In the Minimal Composite Higgs
Models (MCHM) the gauge symmetry $SO(5) \times U(1)_X \times SU(3)$ is
broken down to the SM gauge group on the ultraviolet (UV) boundary and to
$SO(4) \times U(1)_X \times SU(3)$ on the infrared. The Higgs coupling
modifications in these models can be 
described by one single parameter, given by $\xi$. For the fermions,
they depend on the representations of the bulk symmetry into which the
fermions are embedded. In the model MCHM4 based on 
Ref.~\cite{Agashe:2004rs} the fermions transform in the spinorial representation 
of the global symmetry, in the model MCHM5 based on Ref.~\cite{Contino:2006qr} 
the fermions transform in the fundamental representation. In
Table~\ref{tab:coup} we report the modifications of the Higgs
couplings to the SM particles with respect to the corresponding SM
couplings in the SILH set-up and in the MCHM4 and
MCHM5. The last two lines list the novel couplings not present in the
SM, {\it i.e.}~the 2-Higgs-2-fermion coupling and the effective
single and double Higgs couplings to a gluon pair, as defined in the
Feynman rules derived from the SILH Lagrangian,
\beq
hh f\bar{f} &:& -i g_{hhf\bar{f}} \label{eq:coupdef1}\\[0.2cm]
hgg &:& i \delta^{ab} \frac{\alpha_s}{3\pi
 v} [k_1^\nu k_2^\mu - (k_1 \cdot k_2) g^{\mu\nu}]
g_{hgg} \label{eq:coupdef2}\\[0.2cm] 
hhgg &:& i \delta^{ab} \frac{\alpha_s}{3\pi
 v^2} [k_1^\nu k_2^\mu - (k_1 \cdot k_2) g^{\mu\nu}] g_{hhgg} \;
\label{eq:coupdef3}
\eeq
where $k_{1,2}$ denote the incoming momenta of the two gluons $g_\mu^a
(k_1)$ and $g_\nu^b (k_2)$. The effective gluon couplings are not
present in MCHM4 and MCHM5.\s  
\begin{table}[t!]
\begin{center}
\begin{tabular}{|c||c|c|c|}
\hline
& SILH & MCHM4& MCHM5 \\ \hline \hline
$g_{hVV}/g_{hVV}^{\text{SM}}$ & $1 -c_H \,\xi/2$ &
$\sqrt{1-\xi}$ & $\sqrt{1-\xi}$\\
$g_{hf\bar{f}}/g_{hf\bar{f}}^{\text{SM}}$ & $1- (c_H/2
+ c_y)\,\xi$& $\sqrt{1-\xi}$ &  
$\frac{1-2\xi}{\sqrt{1-\xi}}$ \\
$g_{hhh}/g_{hhh}^{\text{SM}}$ & $ 1+(c_6-3c_H/2)\,\xi$& $\sqrt{1-\xi}$ & 
$\frac{1-2\xi}{\sqrt{1-\xi}}$ \\ \hline\hline
$g_{hhf\bar{f}}$& $-(c_H+3c_y)\,\xi \,m_f/v^2$& $-\xi m_f/v^2$ &
$-4\xi \, m_f/v^2$ \\
$g_{hgg}$ and $g_{hhgg}$ & $ 3c_{g}(y_{t}^{2}/g_{\rho}^{2})\xi$ & $0$& $0$ \\
\hline
\end{tabular}
\end{center}
\caption{Higgs couplings to the SM particles (massive gauge bosons
  $V\equiv Z,W$ and fermions) and Higgs self-couplings in the SILH
  set-up, the MCHM4 and MCHM5 normalized to the corresponding
  couplings in the SM, $g_X/g_X^{\text{SM}}$. The last
  two lines summarize the novel couplings not present in the SM, the
  2-Higgs-2-fermion coupling and the effective single and
  double Higgs couplings to a gluon pair as defined in
  Eqs.~(\ref{eq:coupdef1})-(\ref{eq:coupdef3}).} \label{tab:coup} 
\end{table}

In composite Higgs models fermion mass generation can be achieved by the
principle of partial compositeness \cite{partial}. The SM fermions are elementary
particles that couple linearly to heavy states of the strong sector
with equal quantum numbers under the SM gauge group. In particular the
top quark can be largely composite. 
But also the bottom quark can have a sizeable
coupling to heavy bottom partners. For gluon fusion this not only
means that new bottom and top partners are running in the loops
but mixing effects also induce further changes in the top- and
bottom-Higgs Yukawa couplings. In addition to the MCHM4 and 5 models
involving only the pure non-linearities of the Higgs boson in the
Higgs couplings, we consider a model with heavy top and bottom
partners based on the minimal $SO(5)\times U(1)_X/SO(4)\times U(1)_X$
symmetry breaking pattern. The additional $U(1)_X$ is introduced to
guarantee the correct fermion charges. The new fermions transform in the 
antisymmetric representation ${\bf 10}$ of $SO(5)$ in this model
  MCHM10, given by 
\begin{equation}
\begin{split}
&\mathcal{Q}=\frac{1}{2} \times \\
&\begin{pmatrix}0&-(u+u_1)&\frac{i( d-\chi)}{\sqrt{2}}+
\frac{i(d_1-\chi_1)}{\sqrt{2}}&\frac{ d+\chi}{\sqrt{2}}-
\frac{d_1+\chi_1}{\sqrt{2}}&d_4+\chi_4\\
u_1+u&0&\frac{d_1+\chi_1}{\sqrt{2}}+\frac{d+\chi}{\sqrt{2}}&
\frac{i(d_1-\chi_1)}{\sqrt{2}}-\frac{i( d-\chi)}{\sqrt{2}}&-i(d_4-\chi_4)\\
 -\frac{i(d_1-\chi_1)}{\sqrt{2}}-\frac{i( d - \chi)}{\sqrt{2}}&
-\frac{d_1+\chi_1}{\sqrt{2}}-\frac{d + \chi}{\sqrt{2}}
&0&u_1-u&t_4+T_4\\
 \frac{d_1+\chi_1}{\sqrt{2}}-\frac{d + \chi}{\sqrt{2}}&\frac{i(\chi_1-d_1)
} { \sqrt {2}}+\frac{i(d-\chi)}{\sqrt{2}}
&u-u_1&0&-i(t_4-T_4)\\
 -d_4-\chi_4&i(d_4-\chi_4)&-t_4-T_4&i(t_4-T_4)&0\end{pmatrix}\end{split}
\end{equation}
with the electric charge-2/3 fermions $u, u_1, t_4$ and $T_4$, the
fermions $d, d_1$ and $d_4$ with charge $-1/3$, and  the $\chi, \chi_1$
and $\chi_4$ with charge 5/3. The coset $SO(5)/SO(4)$ leads to four
Goldstone bosons, among which three provide the longitudinal modes of
the massive vector bosons $W^\pm$ and $Z$, and the remaining one is the
Higgs boson. The four Goldstone bosons can be parameterized in terms
of the field 
\begin{equation}
 \Sigma=\Sigma_0 \exp(\Pi(x)/f), \hspace*{1cm}
\Sigma_0=(0,0,0,0,1)\;,\hspace*{1cm}\Pi(x)=-i\sqrt{2}
T^{\hat{a}}h^{\hat{a}}(x)\;,
\end{equation}
with the generators $T^{\hat{a}}$ ($\hat{a}=1,...,4$) of the coset $SO(5)/SO(4)$
\begin{equation}
(T^{\hat{a}})_{ij}=-\frac{i}{\sqrt{2}}\left(\delta^{\hat{a}}
_i\delta^5_j-\delta^{\hat{a}}_j\delta^5_i\right)\;.
\end{equation}
The generators of the $SU(2)_{L,R}$ in the fundamental representation read
$(a,b,c=1,2,3, \, i,j=1,...,5)$,
\begin{align}
 (T^{a}_{L/R})_{ij}&=-\frac{i}{2}\left[\frac{1}{2}\epsilon^{abc}
(\delta^b_i\delta^c_j-\delta^b_j\delta^c_i)\pm\delta^a_i\delta^4_j \mp\delta^4_i
\delta ^a_j\right]\;.
\end{align}
The non-linear $\sigma$-model describing the effective low-energy
physics of the strong sector is given by the Lagrangian 
\begin{equation}
\begin{split}
 \mathcal{L}=&\frac{f^2}{2}\left(D_{\mu}\Sigma\right)\left(D^{\mu}
\Sigma\right)^T+\, i\,\text{Tr}( \bar{\mathcal{Q}}_R \slashed{D}\mathcal{Q}_R)+
i\,\text{Tr}(\bar{\mathcal{Q}}_L\slashed{D}\mathcal{Q}_L)\\&+i\bar{q}_{L}
\slashed{D}q_{L} + i\bar{b}_{R}\slashed{D}b_{R}+i\bar{t}_{R}
\slashed{D}t_{R}\\ &-M_{10} \text{Tr}( \bar{\mathcal{Q}}_R \mathcal{Q}_L)-y
 f\left(\Sigma^\dagger \bar{\mathcal{Q}}_R \mathcal{Q}_L \Sigma\right)+h.c.
\\&-\lambda_t \bar{t}_R u_{1L}-\lambda_b \bar{b}_R d_{1L}
-\lambda_q (\bar{T}_{4R}, \bar{d}_{4R})q_L+h.c.\label{lagrangian}\;,
\end{split}
\end{equation}
with the covariant derivative 
\begin{equation}
D_{\mu}\Sigma=\partial_{\mu}\Sigma -i g' B_{\mu} \Sigma(T^3_R+X)-i g W_{\mu}^a
\Sigma T^a_L
\end{equation}
in terms of the $SU(2)_L$ and $U(1)_Y$ gauge fields $W_\mu^a$ and
$B_\mu$, respectively, with their corresponding couplings $g$ and
$g'$. 
The bilinear terms in the fermion fields lead to mass matrices for the 2/3, 
$-1/3$ and 5/3 charged fermions, when the Higgs field is shifted by its
VEV $\langle H \rangle$, $H = \langle H\rangle + h$. The mass
matrices can be diagonalized by 
means of a bi-unitary transformation. The  2-fermion couplings to one
and two Higgs bosons are obtained by expanding  
the mass matrices in the interaction eigenstates up to first,
respectively, second order in the Higgs field, and subsequent
transformation into the mass eigenstate basis. The mass matrices and
the coupling matrices of one Higgs boson to two bottom-like and top-like
states can be found in Ref.~\cite{Gillioz:2013pba}. In the
Appendix~\ref{app:coupmass} we give the coupling matrices for the
2-Higgs-2-fermion couplings and, for completeness, repeat the
matrices given in Ref.~\cite{Gillioz:2013pba}.

\section{Next-to-leading Order QCD Corrections to Higgs Pair Production in Composite Higgs Models \label{sec:NLOQCD}}
The NLO QCD corrections to Higgs pair production
in the SM have been computed in~Ref.~\cite{Dawson:1998py} by applying the
heavy top approximation, in which the heavy fermion loops are 
replaced by effective vertices of gluons to Higgs bosons. These can be
obtained by means of the low-energy theorem (LET) \cite{let1,let2}. The Higgs field 
is treated here as a background field, and the field-dependent mass of
each heavy particle is taken into account in the gluon
self-interactions at higher orders.
The LET provides the zeroth order in an expansion in
small external momenta. Since in Higgs pair production the
requirements for such an expansion are not fulfilled sufficiently
  reliably, it fails to give 
accurate results for the cross section at LO \cite{Glover:1987nx}. In the context of 
composite Higgs models, the discrepancy between the LO cross section with 
full top quark mass dependence and the LO cross section in the LET approximation 
is even worse \cite{Gillioz:2012se}. For relative higher order corrections the LET
approximation should, however, become better, if the LO order cross section is 
taken into account with full mass dependence. This is because the dominant 
corrections given by the soft and collinear gluon corrections factorize
from the LO cross section generating a part independent of the
masses of the heavy loop particles relative to the LO cross section.  
This was confirmed in Ref.~\cite{Grigo:2013rya} by including higher terms in 
the expansion of the cross section in small external momenta. Based on
these findings, in this section we will give the NLO QCD corrections
for Higgs pair production in composite Higgs models in the LET
approach. \s

\begin{figure}[b!]
\includegraphics[width=0.3\linewidth,angle=-90]{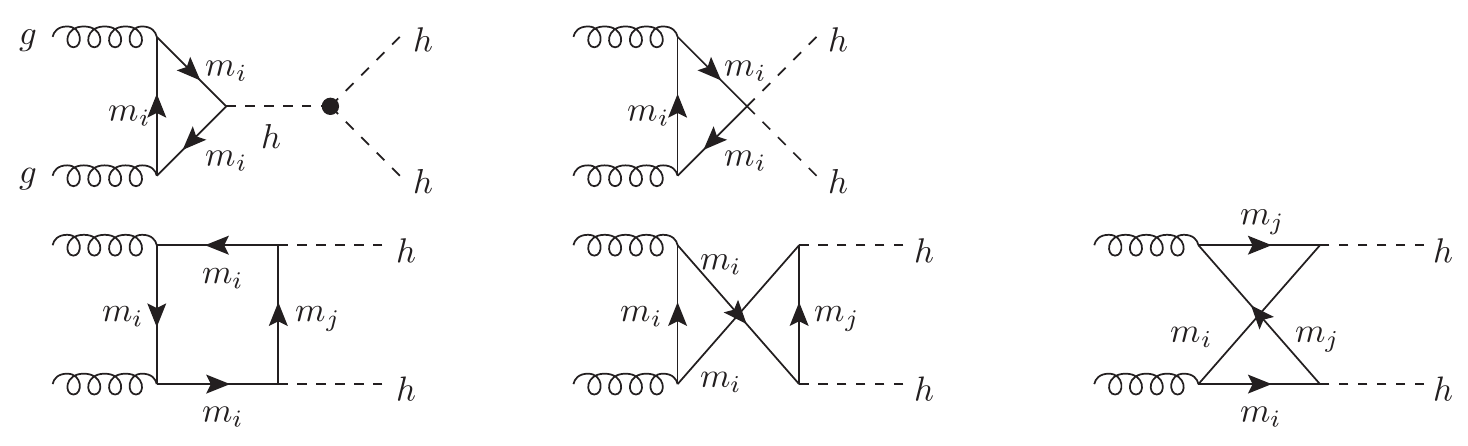}
 \caption{Generic Feynman diagrams contributing to $gg\to 
hh$ at LO, with $n$ novel fermionic resonances, the top and the
bottom quark with masses $m_i$ ($i=1,...,n,t,b$). The index $j$ is
introduced to indicate where different fermions can contribute in the
loop. \label{fig:gglodiags}} 
\end{figure}

The expression of the LO gluon fusion into Higgs pairs in a composite
Higgs model with heavy top partners has been given in
\cite{Gillioz:2012se}. It can be taken over here, by simply extending
the sum to include also the bottom quark and its partners. We
summarize here the most important features and refer to
\cite{Gillioz:2012se} for more details. The generic diagrams that contribute
to the process at LO are depicted in Fig.~\ref{fig:gglodiags}.
Besides the new 2-Higgs-2-fermion coupling $f\bar{f}hh$ the
additional top and bottom partners in the loops have to be taken into
account. These lead also to new box diagrams involving off-diagonal Yukawa
couplings, with, respectively, the top and its heavy charge-2/3
partners or the bottom and its heavy partners of charge $-1/3$. The
hadronic cross section is obtained by convolution with the parton
distribution functions $f_g$ of the gluon in the proton,
\beq
\sigma (pp \to hh + X) = \int_{\tau_0}^1 d\tau \int_\tau^1 \frac{dx}{x} \,
f_g(x,\mu_F) \, f_g(\tau/x,\mu_F) \, \hat{\sigma}_{\text{LO}} (\tau s) \;,
\eeq
where $s$ denotes the squared hadronic c.m.~energy, $\mu_F$ the
factorization scale and
\beq
\tau_0 = \frac{4 m_h^2}{s} 
\eeq
in terms of the Higgs boson mass $m_h$.
The partonic LO cross section can be cast into the form
\beq
\hat{\sigma}_{\text{LO}} (gg\to hh) \!\!&=&\!\! \frac{\alpha_s^2 (\mu_R)}{1024
  (2\pi)^3 \hat{s}^2} \int_{\hat{t}_-}^{\hat{t}_+} d\hat{t} \left[
\Big| \sum_{q=t,b}\left(
    F_{\Delta}^{\text{LO}}+F_{\Box}^{\text{LO}}\right)\Big|^2+\Big|
  \sum_{q=t,b} G_{\Box}^{\text{LO}} \Big|^2 \right]\;,
\eeq
with the strong coupling constant $\alpha_s$ at the renormalization
scale $\mu_R$. We have introduced the Mandelstam variables 
\beq
\hat{s} =\tau s = Q^2 \;, \quad \hat{t} = m_h^2 - \frac{Q^2 (1-\beta
  \cos\theta)}{2} \quad \mbox{and} \quad
\hat{u} = m_h^2 - \frac{Q^2 (1+\beta \cos\theta)}{2}  
\;,
\eeq
in terms of the scattering angle $\theta$ in the partonic c.m.~system
with the invariant Higgs pair mass $Q$ and the relative velocity
\beq
\beta = \sqrt{1-\frac{4m_h^2}{Q^2}} \;.
\eeq
The integration limits at $\cos\theta = \pm 1$ are given by
\beq
\hat{t}_\pm = m_h^2 - \frac{Q^2 (1\mp \beta)}{2} \;.
\eeq
The form factors read
\begin{align}
F_{\Delta}^{\text{LO}}&=\sum_{i=1}^{n_q}C_{i,\Delta} F_\Delta (m_i) \label{eq:ff1} \\
F_{\Box}^{\text{LO}}&= \sum_{i=1}^{n_q} \sum_{j=1}^{n_q} \left(
  g_{h\bar{q}_i q_j}^2 F_\Box (m_i,m_j) + g_{h\bar{q}_i q_j,5}^2
  F_{\Box,5} (m_i,m_j) \right) \label{eq:ff2}\\ 
G_{\Box}^{\text{LO}}&=\sum_{i=1}^{n_q} \sum_{j=1}^{n_q} \left(
    g_{h\bar{q}_i q_j}^2 G_\Box (m_i,m_j) + g^2_{h\bar{q}_i q_j,5}
    G_{\Box,5} (m_i,m_j) \right)\;. \label{eq:ff3}
\end{align}
The triangle and box form factors $F_\Delta$,
$F_\Box$, $F_{\Box,5}$, $G_\Box$ and $G_{\Box,5}$ can be found in the
appendices of~\cite{Gillioz:2012se,PhD}.\footnote{The form factors
  $F_\Delta, F_\Box$ and $G_\Box$ relate to those given in
  Ref.~\cite{Plehn:1996wb} for the SM case as $F_\Delta (m)= \hat{s}/m \,
  F_\Delta^{\text{SM}} (m)$, $F_\Box= \hat{s}/m^2
  F_\Box^{\text{SM}} (m)$ and $G_\Box (m) = \hat{s}/m^2
  G_\Box^{\text{SM}} (m)$.}
The sum runs up to $n_t=5$ for the top quark and its charge-2/3 partners and up
to $n_b=4$ in the bottom sector. The couplings are defined as
\beq
g_{h\bar{q}_i q_j} = \frac{1}{2} ( G_{hqq,ij} + G_{hqq,ji} ) \; ,
\qquad
g_{h\bar{q}_i q_j,5} = \frac{1}{2} ( G_{hqq,ji} - G_{hqq,ij} ) 
\eeq
and
\beq
g_{hh\bar{q}_i q_j} = ( G_{hhqq,ij} + G_{hhqq,ji} ) \; ,
\eeq
where $G_{hqq,ij}$ and $G_{hhqq,ij}$ denote the ($i$th,$j$th) matrix
elements of the coupling matrices in Eq.~(\ref{eq:tbcoups}) of
the appendix. The triangle factor $C_{i,\Delta}$ reads
in the MCHM10
\beq
C_{i,\Delta} = \frac{g_{hhh} g_{h\bar{q}_i
    q_i}}{Q^2-m_h^2+im_h\Gamma_h} + g_{hh\bar{q}_i q_i} \quad
\mbox{with} \quad  
g_{hhh} = \frac{3m_h^2}{v} \frac{1-2\xi}{\sqrt{1-\xi}} \;,
\eeq
as given in the MCHM5. In the SM and in the
composite Higgs models MCHM4 and MCHM5 involving solely the Higgs
non-linearities and no heavy fermionic resonances, no sum over 
heavy top and bottom partners contributes and only a sum over the top
and bottom running in the loop has to be performed, {\it i.e.}~$n_t=n_b=1$, with
$m_i=m_j=m_q$ and $q=t,b$, and hence also 
\beq
g_{h\bar{q}_i q_j,5}=0 \qquad \mbox{for SM, MCHM4 and MCHM5.}
\label{eq:coups1}
\eeq
The Yukawa couplings read 
\beq
g_{h\bar{q} q}^{\text{SM}} = \frac{m_q}{v} \;, \quad
g_{h\bar{q}q}^{\text{MCHM4}} = g_{h\bar{q} q}^{\text{SM}} \, \sqrt{1-\xi} \qquad
\mbox{and}  \qquad
g_{h\bar{q}q}^{\text{MCHM5}} = g_{h\bar{q} q}^{\text{SM}} \,
\frac{1-2\xi}{\sqrt{1-\xi}}  \;, \label{eq:coups2}
\eeq
and for the 2-Higgs-2-fermion coupling we have 
\beq
g_{hh\bar{q} q}=0 \qquad \mbox{in the SM and} \qquad
g_{hh\bar{q} q}^{\text{MCHM4}} = -\xi \, \frac{m_q}{v^2} \quad \mbox{and}
\quad
g_{hh\bar{q} q}^{\text{MCHM5}} = -4 \xi \, \frac{m_q}{v^2} \;,
\label{eq:coups3}
\eeq
while the Higgs self-coupling becomes
\beq
g_{hhh}^{\text{SM}} = \frac{3m_h^2}{v} \;, \quad
g_{hhh}^{\text{MCHM4}} = g_{hhh}^{\text{SM}} \, \sqrt{1-\xi} \qquad
\mbox{and}  \qquad
g_{hhh}^{\text{MCHM5}} = g_{hhh}^{\text{SM}} \,
\frac{1-2\xi}{\sqrt{1-\xi}} \;.
\label{eq:coups4}
\eeq

\begin{figure}[h!]\vspace*{-2cm}
\hspace*{-1.5cm}
\includegraphics[width=0.9\linewidth]{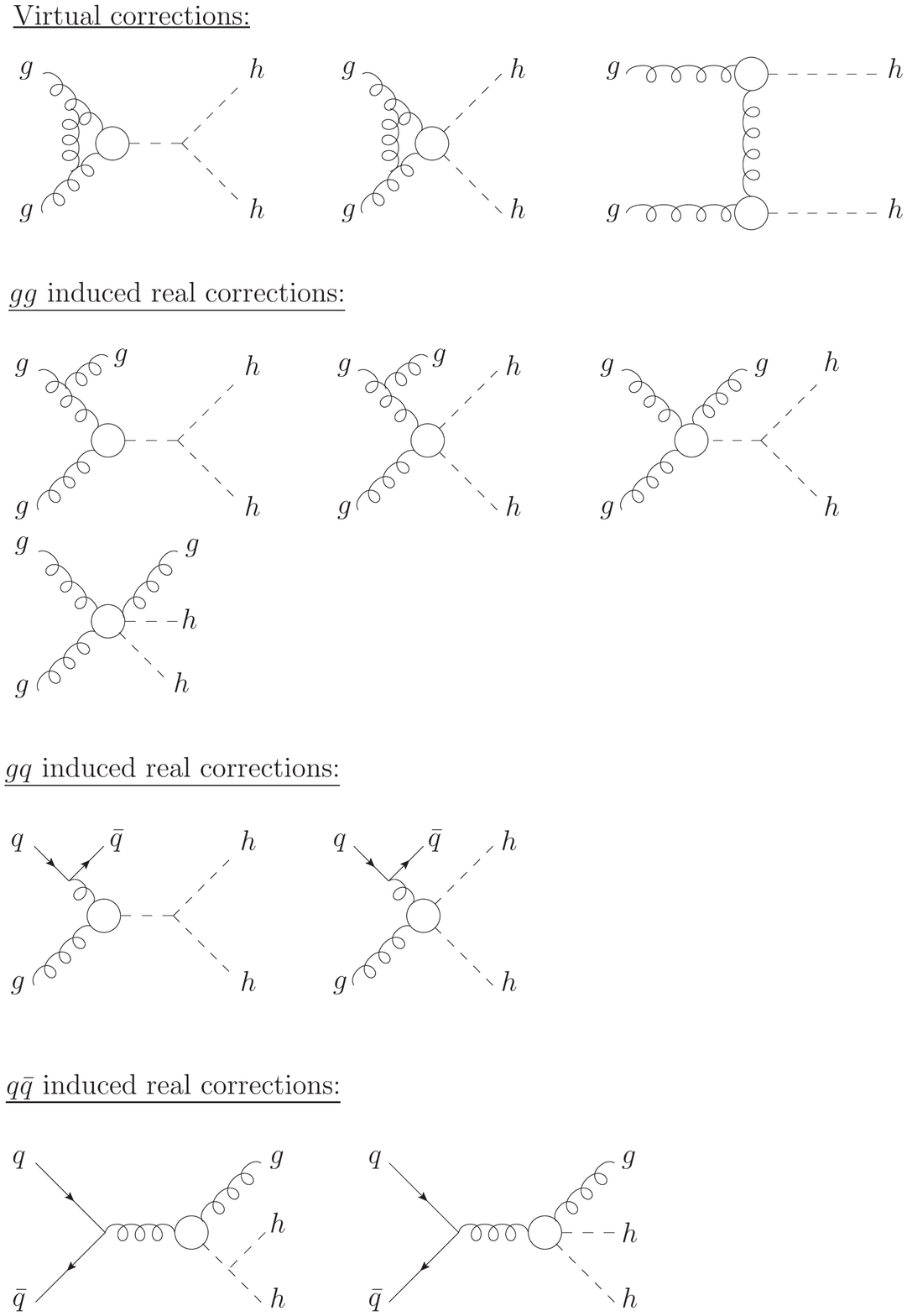}\vspace*{-2.5cm}
 \caption{Generic Feynman diagrams contributing to the NLO QCD
   corrections to $gg\to hh$.\label{fig:CHM8}}
\end{figure}

The Feynman diagrams contributing to Higgs pair production at NLO QCD
are shown in Fig.~\ref{fig:CHM8}. The blob in the figure marks the
effective vertices of gluons to Higgs boson(s). The first three Feynman
diagrams show the virtual contributions. The remaining Feynman
diagrams of Fig.~\ref{fig:CHM8} display the real corrections
generically ordered by the initial states $gg$, $gq$ and $q\bar{q}$. 
At NLO the cross section is then given by
\begin{equation} 
\sigma_{\text{NLO}} (pp \to hh + X)
=\sigma_{\text{LO}}+\Delta\sigma_{\text{virt}}+\Delta\sigma_{gg} 
+\Delta\sigma_{gq}+\Delta\sigma_{q\bar{q}}\,.\label{CHM33}
\end{equation}
The individual contributions in Eq.~(\ref{CHM33}) read
\beq
\sigma_{\text{LO}} &=& \int_{\tau_0}^1 d\tau \, \frac{d {\cal
    L}^{gg}}{d\tau} \, \hat{\sigma}_{\text{LO}} (Q^2 = \tau s) \\
\Delta \sigma_{\text{virt}} &=& \frac{\alpha_s (\mu_R)}{\pi} \int_{\tau_0}^1
d\tau \, \frac{d {\cal L}^{gg}}{d\tau} \, \hat{\sigma}_{\text{LO}} (Q^2 =
\tau s) \, C \label{eq:contrib1}\\
\Delta \sigma_{gg} &=& \frac{\alpha_s (\mu_R)}{\pi} \int_{\tau_0}^1
d\tau \, \frac{d {\cal L}^{gg}}{d\tau} 
\int_{\tau_0/\tau}^1 \frac{dz}{z} \, \hat{\sigma}_{\text{LO}} (Q^2 =
z\tau s) \left\{ -z P_{gg} (z) \log \frac{\mu_F^2}{\tau s} \right. \nonumber \\
&& \hspace*{2cm} \left. -\frac{11}{2} (1-z)^3 + 6[1+z^4+(1-z)^4] \left(
    \frac{\log (1-z)}{1-z} \right)_+ \right\} \label{eq:contrib2} \\
\Delta \sigma_{gq} &=& \frac{\alpha_s (\mu_R)}{\pi} \int_{\tau_0}^1
d\tau \, \sum_{q,\bar{q}} \frac{d {\cal L}^{gq}}{d\tau} 
\int_{\tau_0/\tau}^1 \frac{dz}{z} \, \hat{\sigma}_{\text{LO}} (Q^2 =
z\tau s) \left\{ - \frac{z}{2} P_{gq} (z) \log \frac{\mu_F^2}{\tau s
    (1-z)^2} \right. \nonumber \\
&& \left. \hspace*{5cm} + \frac{2}{3} z^2 - (1-z)^2
\right\} \label{eq:contrib3} \\
\Delta \sigma_{q\bar{q}} &=& \frac{\alpha_s (\mu_R)}{\pi} \int_{\tau_0}^1
d\tau \, \sum_q \frac{d {\cal L}^{q\bar{q}}}{d\tau} 
\int_{\tau_0/\tau}^1 \frac{dz}{z} \, \hat{\sigma}_{\text{LO}} (Q^2 =
z\tau s) \, \frac{32}{27} (1-z)^3 \;, \label{eq:contrib4}
\eeq
with the Altarelli-Parisi splitting functions given by \cite{altaparisi}
\beq
P_{gg} (z) &=& 6 \left\{ \left( \frac{1}{1-z} \right)_+ + \frac{1}{z}
  - 2 + z (1-z) \right\} + \frac{33-2N_F}{6} \delta (1-z) \nonumber \\
P_{gq} (z) &=& \frac{4}{3} \frac{1+(1-z)^2}{z} \;,
\eeq
and $N_F=5$ in our case.
The real corrections $\Delta\sigma_{gg}$, $\Delta\sigma_{gq}$ and
$\Delta\sigma_{q\bar{q}}$ have straightforwardly been obtained from
Ref.~\cite{Dawson:1998py} by replacing the LO cross section of the SM
with the LO cross section for composite Higgs models. The calculation of
$\Delta \sigma_{\text{virt}}$ is a bit more involved. While the first two
diagrams factorize from the LO cross section and can hence directly be
taken over from the SM, the third diagram in Fig.~\ref{fig:CHM8} does
not factorize and needs to be recalculated for the composite Higgs
case. 
The virtual coefficient $C$ is then found to be
\begin{equation}
\begin{split}
C= &\pi^2 + \frac{11}{2} + \frac{33-2N_F}{6} \log \frac{\mu_R^2}{Q^2} \\
+&\text{Re}\frac{ 
\int_{\hat t_-}^{\hat t_+} d\hat t \, \frac{4}{9} (g_{hgg}^{\text{eff}})^2
\, \biggl\{\left(F_{\Delta}^{\text{LO}}+F_{\Box}^{\text{LO}}\right) -  \frac{p_T^2}{2 \hat t 
\hat u} (Q^2-2 m_h^2) \, G_{\Box}^{\text{LO}}\biggr\}} 
{\int_{\hat t_-}^{\hat t_+} d\hat t \,   \left[\Big|    \sum_{q=t,b} \left(  
F_{\Delta}^{\text{LO}}+F_{\Box}^{\text{LO}} \right)\Big|^2
+ \Big|  \sum_{f=t,b}  G_{\Box}^{\text{LO}}\Big|^2 \right]}
\label{CHM36}
\end{split}
\end{equation}
with
\begin{equation}
 p_T^2=\frac{(\hat{t}-m_h^2)(\hat{u}-m_h^2)}{Q^2}-m_h^2\,.
\end{equation}
The first line in Eq.~\eqref{CHM36} corresponds to the NLO
  contribution from the first two diagrams in Fig.~\ref{fig:CHM8},
while the second line corresponds to the
  NLO contribution from the third diagram of Fig.~\ref{fig:CHM8}. 
The factor $(g_{hgg}^{\text{eff}})^2$ stems from the two effective
  Higgs-gluon-gluon vertices in diagram 3 of Fig.~\ref{fig:CHM8}. This
  vertex is obtained by integrating out all heavy loop particles in
  the loop-induced Higgs coupling to gluons defined in 
  Eq.~(\ref{eq:coupdef2}) with $g_{hgg} \equiv g_{hgg}^{\text{eff}}$ and
\begin{equation}
g_{hgg}^{\text{eff}}=\left(\sum_{i=1}^{n_t}\frac{g_{h\bar{q}_iq_i}v}{m_{i}}+
 \sum_{i=1}^{n_{\tilde{b}}}
 \frac{g_{h\bar{q}_iq_i}v}{m_i}
\right)\,.\label{CHM37}
\end{equation}
The first term is the sum over the normalized top quark and top
partner couplings and the second term the sum over the normalized
bottom partner couplings to the Higgs boson, excluding consistently
the light bottom quark contribution from the loop. 
The composite Higgs cross sections for MCHM4, MCHM5 and for the
composite Higgs model with heavy top and bottom partners, including the NLO
corrections have been implemented in {\tt HPAIR}~\cite{HPAIR}. In order
to exemplify the impact of the NLO QCD corrections, we consider the
simple case with the pure Higgs non-linearities only and the fermions
transforming in the fundamental representation, {\it i.e.} the
benchmark model MCHM5, see Table~\ref{tab:coup}. 
The coupling $g_{hgg}^{\text{eff}}$ then reduces to
$g_{hgg}^{\text{MCHM5}}  = (1-2\xi)/\sqrt{1-\xi}$ and the remaining
couplings are given in Eqs.~(\ref{eq:coups1})-(\ref{eq:coups4}). 
\begin{figure}
\centering
\vspace*{-0.5cm}
\includegraphics[width=0.7\linewidth, 
angle=0]{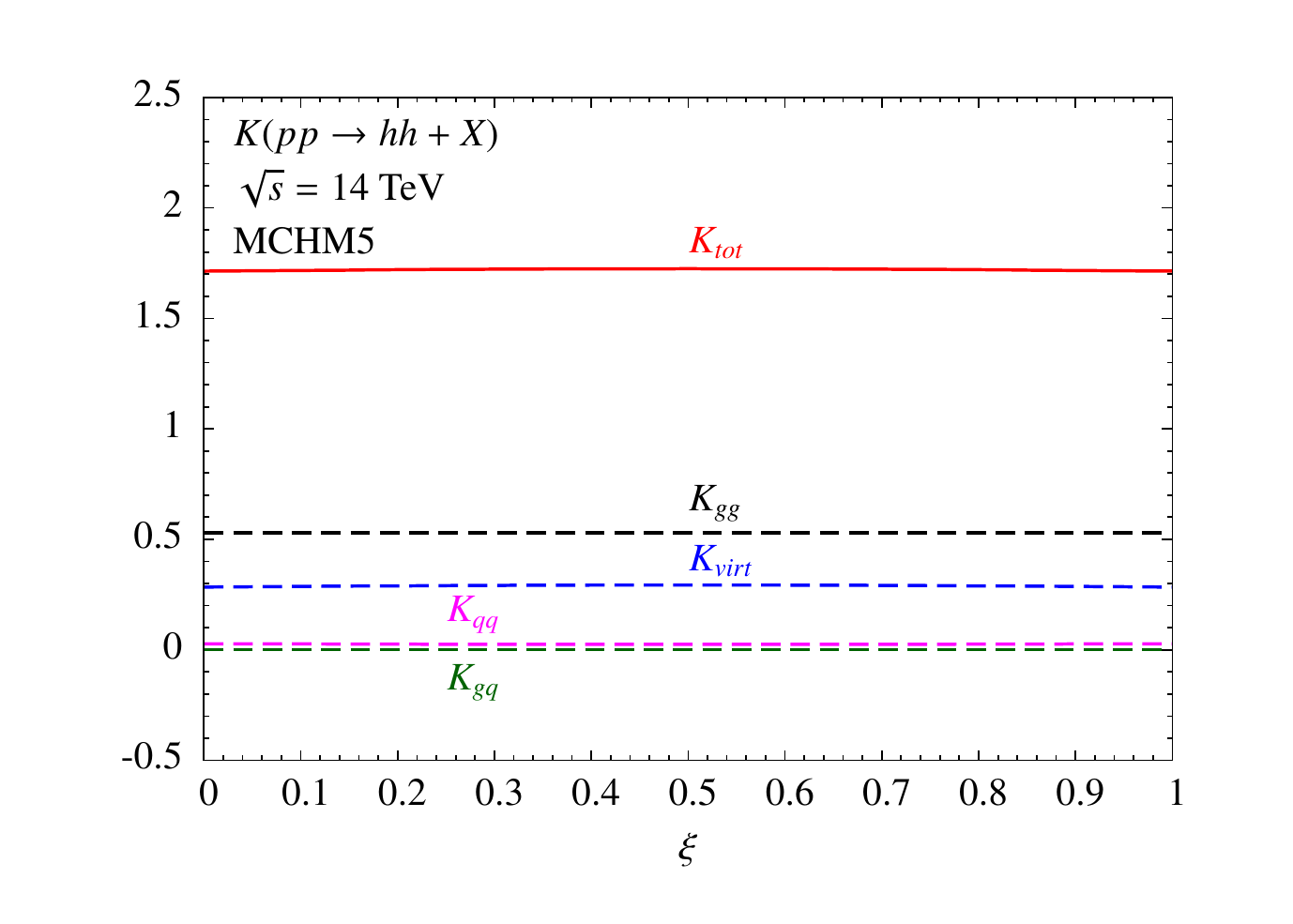}\vspace*{-0.5cm}
 \caption{$K$-factors for the $pp\to hh+X$ cross section with
   $m_h=125$~GeV in MCHM5 
   as a function of $\xi$ for the scale choice
   $\mu_F=\mu_R=m_{hh}/2$ and the 
c.m.~energy $\sqrt{s}=14$~TeV. The dashed
lines show the individual contributions of the virtual contributions
and the real corrections according to their initial states. \label{fig:CHM9}} 
\end{figure}
We define the $K$-factors for the total cross section and the
individual contributions as
\begin{equation}
K=\frac{\sigma_{NLO}}{\sigma_{LO}} \qquad \mbox{and} \qquad
K_i=\frac{\Delta\sigma_{i}}{\sigma_{LO}} \;, \qquad i=\text{virt}, gg, gq,
q\bar{q} \;.
\end{equation}
The cross section at LO is computed with the full quark mass 
dependences. As the NLO cross section in the LET approximation only
includes top quark contributions\footnote{Note, that in MCHM5 we
    have no heavy top or bottom partners.}, at LO we consistently neglect also
the bottom quark contributions, which in the SM amount to 1\% for a
125~GeV Higgs boson. The c.m.~energy $\sqrt{s}=14$~TeV
  and the top and bottom quark mass are set to 
$m_t=173.2$~GeV and $m_b=4.75$~GeV, respectively. The cross
sections are computed with the MSTW2008 PDF set \cite{MSTW}. The
strong coupling constant is evaluated at the corresponding loop order with 
\beq
\alpha_s^{\text{LO}}(M_Z)=0.13939 \qquad \mbox{and} \qquad
\alpha_s^{\text{NLO}}(M_Z)=0.12018 \;. 
\eeq
The renormalization and factorization scales are set
to $\mu_R=\mu_F=m_{hh}/2$, where $m_{hh}$ denotes the invariant
Higgs pair mass. Figure \ref{fig:CHM9} displays the 
results for the $K$-factors for the MCHM5 as a function of  
$\xi$. The solid line shows the total $K$-factor, the dashed lines are the 
individual contributions. As can be inferred from the plot, the real
and virtual corrections of the $gg$ initial state make up the 
bulk of the QCD corrections. The $qg$ and the
$q\bar{q}$ initiated real radiation diagrams only lead to a small correction.
The $K$-factor is almost independent of $\xi$. In the real corrections, the 
Born cross  section, which shows the only dependence on $\xi$,
almost completely drops out numerically. For the virtual
contributions some dependence on $\xi$ may be expected. The virtual
correction due to the constant term in $C$, {\it i.e.}~the first line
in Eq.~(\ref{CHM36}) does not develop any dependence on
$\xi$, however, as it factorizes from the LO cross
section. The dependence of $\xi$ can only emerge from the second line
in Eq.~\eqref{CHM36}, which, however, is numerically suppressed.
This is already the case in the SM, where the
corresponding term contributes less than 3\% to
$\Delta\sigma_{\text{virt}}$. This result also holds true for the case were the
heavy quark partners are explicitly included. In
composite Higgs models, the NLO QCD corrections to Higgs pair
production can hence well be approximated by
multiplying the full LO cross section of the composite Higgs model
under consideration with the SM $K$-factor. \s

Figure \ref{fig:CHM9} can also be obtained by using the
results of Ref.~\cite{Grober:2015cwa}. Note however, that the effects
of heavy top and bottom partners in the effective field theory
computation of Ref.~\cite{Grober:2015cwa} have to be added to the top
quark contribution, encoded into the Wilson coefficients in front of
the operators $hG^{\mu \nu}G_{\mu \nu}$ and $hhG^{\mu \nu}G_{\mu \nu}$.

\section{Numerical Analysis of New Physics Effects in Higgs Pair Production via Gluon Fusion \label{sec:numerical1}}
Having derived the NLO QCD corrections, we can now turn to the analysis of  
NP effects in Higgs pair production. We assume that no NP is found 
before Higgs pair production becomes accessible. This means that we
require deviations in the Higgs boson couplings with respect to
the SM to be smaller than the projected sensitivities of
the coupling measurements at  
an integrated luminosity of $300\;\text{fb}^{-1}$ and
$3000\;\text{fb}^{-1}$, respectively. For the projected sensitivities we take
the numbers reported in Ref.~\cite{Englert:2014uua}. Similar numbers
can be found in Refs.~\cite{SFitter}. 
In our analysis we focus on the most promising final states, given
  by $b\bar{b}\gamma\gamma$ and 
$b\bar{b}\tau^+ \tau^-$
\cite{Baglio:2012np,Baur:2003gpa,Baur:2002rb,Baur:2003gp}. \s 

We call Higgs pair production to be sensitive to NP if the difference 
between the number of signal events $S$ in the considered NP model
and the corresponding number $S_{SM}$ in the SM exceeds a minimum of
3 statistical standard deviations, {\it i.e.}
\begin{equation}
 S_{SM}+\beta\sqrt{S_{SM}}\leq S\hspace*{1cm}\text{or}\hspace*{1cm}  
S_{SM}-\beta\sqrt{S_{SM}}\geq S \label{crit}
\end{equation}
with $\beta=3$ for a $3\sigma$ deviation. The signal events are obtained 
as 
\beq
S=\sigma\cdot BR \cdot L\cdot A \;,
\eeq
where $BR$ denotes the branching ratio into the respective final states,
$L$ the integrated luminosity and $A$ the
acceptance due to cuts 
applied on the cross section. The acceptances have been extracted from  
Ref.~\cite{Baglio:2012np} for the $b\bar{b}\gamma\gamma$ and 
$b\bar{b} \tau^+ \tau^-$ final states. The acceptance for the BSM
signal only changes slightly, as we explicitly checked. \s

In specific models, the correlations of the couplings will lead to stronger 
bounds on the parameters. In particular in the MCHM4 and MCHM5 as introduced in 
section~\ref{sec:chm}, the only new parameter is $\xi$. The value of $\xi$ can 
hence strongly be restricted by Higgs coupling
measurements~\cite{Bock:2010nz}. 
Based on the data for a c.m.~energy of $\sqrt{s}=7$ and 8 TeV at
  an integrated luminosity of 4.6-4.8 fb$^{-1}$ and 20.3 fb$^{-1}$,
  respectively, ATLAS observes a 95\% C.L.~upper limit of $\xi<0.12$ 
  in the MCHM4 and of $\xi<0.15$ in the MCHM5 \cite{ATLASHiggs}. This
restricts the gluon fusion cross section for the 
MCHM4 to the range $40.06\;\text{fb}\leq
\sigma_{\text{MCHM4}}\leq 44.55 
\;\text{fb}$ and for the MCHM5 to $40.06\;\text{fb}\leq
\sigma_{\text{MCHM5}}\leq 89.26 \;\text{fb}$. The value
$\sigma=40.06\;\text{fb}$ corresponds to the SM cross section at NLO  
QCD for $m_t=173.2\;\text{GeV}$ and hence the MCHM4/MCHM5 value 
for $\xi=0$. For the high luminosity options at the LHC we apply the
projected estimates for the parameter $\xi$ given in Ref.~\cite{Englert:2014uua},
\beq
\begin{array}{lll}
\int {\cal L} = 300 \mbox{ fb}^{-1}: & \mbox{MCHM4:} & 0.076\\
& \mbox{MCHM5:} & 0.068 \\[0.1cm]
\int {\cal L} = 3000 \mbox{ fb}^{-1}: & \mbox{MCHM4:} & 0.051\\
& \mbox{MCHM5:} & 0.015
\end{array}
\eeq
Based on these estimates, we give in Table~\ref{tab:cxn} the maximal values for
the cross section times branching ratio. In the fourth and sixth
columns we report whether the process within MCHM4,
respectively, MCHM5 can be distinguished from the SM cross section by
more than $3\sigma$ according to the criteria given in
Eq.~\eqref{crit} for $\beta=3$. In the check of Eq.~\eqref{crit} we
took into account the slight change in the 
acceptance of the signal rate for the composite Higgs models. Due to
the coupling modifications and the new diagram from the
2-Higgs-2-fermion coupling the applied cuts in the analysis of
Ref.~\cite{Baglio:2012np} affect the cross section in a slightly different way.
\s
\begin{table}
\centering
  \begin{tabular}{|cc|cccc|}\hline
  \multicolumn{2}{|c|}{}  & $\sigma_{b\bar{b}\gamma\gamma}$ [fb] & 
$\Delta_{3\sigma}$ &$\sigma_{b\bar{b}\tau^+ \tau^-}$ [fb]& $\Delta_{3\sigma}$ \\ \hline
&$\xi=0.12$  (LHC20.3)&0.119&no&
 3.26&no\\
\multirow{1}{*}{MCHM4}&$\xi=0.076$ (LHC300)&$0.114$&no&
 3.13&no\\
&$\xi=0.051$ (LHC3000)&0.112&no&
3.07&no\\ \hline
 &$\xi=0.15$  (LHC20.3)&0.315&yes&
 5.35&yes\\
\multirow{1}{*}{MCHM5}&$\xi=0.068$ (LHC300)&$0.175$&no&
 3.96&no\\
&$\xi=0.015$ (LHC3000)&$0.119$&no&
 3.14&no\\ \hline
\end{tabular}
\caption{Values of the cross section times branching ratio in the MCHM4 and 
MCHM5 for the maximal allowed values of $\xi$ at $95\%$ C.L. \cite{ATLASHiggs} 
and for the projected values at $\int \mathcal{L}=300\;\text{fb}^{-1}$ and 
$\int \mathcal{L}=3000\;\text{fb}^{-1}$ of
Ref.~\cite{Englert:2014uua}. The fourth and sixth columns decide whether
the Higgs production cross section will develop a 
deviation to the SM Higgs pair production cross section of more than $3 
\sigma$ according to the criteria of Eq.~(\ref{crit}).}\label{tab:cxn}
\end{table}

The table shows, that with the projected precision on $\xi$ at
  high luminosities Higgs pair production in both MCHM4 and MCHM5 leads to
cross sections too close to the SM value to be distinguishable
from the SM case. Although with the present bounds on $\xi$ Higgs pair 
production in MCHM5 differs by more than 3$\sigma$ from the SM
prediction, the corresponding cross section is too small to be
measurable, so that first signs of NP through this process are precluded.

\section{Numerical Analysis for MCHM10
\label{sec:numerical2}}
We consider the MCHM10 with one multiplet of fermionic resonances
below the cut-off. In this model, with more than one parameter
determining the Higgs coupling modifications, 
there is more freedom and a larger allowed parameter space (see
Ref.~\cite{Gillioz:2013pba} for a thorough analysis). This implies,
that the sensitivity on the Higgs couplings is less
constrained. The numbers of the projected sensitivities are taken
from Table I in Ref.~\cite{Englert:2014uua}. Additionally, we  
need to take into account the bounds from the direct searches for
  new fermions. Currently, exotic new fermions with
charge 5/3 are excluded up to masses of $m_{\chi}\le
840\;\text{GeV}$~\cite{Aad:2015mba}, bottom  
partners up to masses of $m_B\le 900\;\text{GeV}$ \cite{Khachatryan:2015gza}
and top partners with masses of $m_T\le 950\;\text{GeV}$ 
\cite{Aad:2015kqa}. Note that the 
latter two limits on the masses depend on the branching ratios of the
bottom and top partner, respectively. These limits are based on  
pair production of the new heavy fermions. First studies for 
single production of a new vector-like fermion were performed in
Refs.~\cite{ATLAS-CONF-2014-036}  
 and can potentially be more important at large energies 
\cite{Aguilar-Saavedra:2013qpa} but are more model-dependent. Due to 
this model dependence it is difficult to estimate the LHC reach on single production
for our case. Hence we will only use the estimated reach on new
vector-like fermions  
in pair production. In Refs.~\cite{Bhattacharya:2013iea, CMS:2013xfa} the 
potential reach of the LHC for charged-2/3 fermions, depending on
  their branching ratios is estimated. Following \cite{CMS:2013xfa} we use the reach 
$m_T\lesssim 1.3 \text{ TeV}$ for $\int \mathcal L=300\text{ fb}^{-1}$ and 
$m_T\lesssim 1.5 \text{ TeV}$ for $\int \mathcal L=3000\text{ fb}^{-1}$.
The potential reach for bottom partners is $m_B\lesssim 1\text{ TeV}$
for $\int \mathcal L=300\text{ fb}^{-1}$ and $m_B \lesssim 1.5 \text{
  TeV}$ for $\int \mathcal L=3000\text{ fb}^{-1}$
\cite{Varnes:2013pxa}. We estimate the additional sensitivity for the
reach of exotic new fermions by multiplying the excluded
  cross section at $\sqrt{s}=8$~TeV with \cite{Bharucha:2013epa}
\begin{equation}
r = \sqrt{\frac{\sigma_{BKG}(14
\text{ TeV})}{\sigma_{BKG}(8\text{ TeV})}
\frac{L_{LHC8}}{L_{LHC14}}
} \;,
\end{equation}
where $L_{\text{LHC8}}$ and $L_{\text{LHC14}}$ denote the integrated
luminosities of the LHC run at $\sqrt{s}=8$ and 14~TeV,
respectively. This implies a reach of $m_{\chi}\approx 1370\text{
  GeV}$ at $\int \mathcal  L=300\text{ fb}^{-1}$ and
$m_{\chi}\approx 1550\text{ GeV}$ at $\int \mathcal L=3000\text{ fb}^{-1}$. For 
the background estimate we only considered the dominant background 
$t\bar{t}W^{\pm}$ \cite{CMSExotic}. The background cross section was computed 
with {\tt MadGraph5} \cite{Alwall:2011uj}. Although the assumption of
stronger projections on the reach of new fermion masses of up to 2 TeV
\cite{Delaunay:2013pwa} will lead to a reduced number of points
allowed by the constraints we are imposing, it will not change our
final conclusion, as we checked explicitly. Note also
that in composite Higgs models there is a connection between the Higgs
boson mass and the fermionic resonances
\cite{Matsedonskyi:2012ym,Pomarol:2012qf}. Reference
\cite{Pomarol:2012qf} finds that the mass of the lightest
 top partner $m_{T_\text{lightest}}$ should be lighter than
\begin{equation}
 m_{T_\text{lightest}}\lesssim \frac{m_h \pi v}{m_t \sqrt{N_c}
   \sqrt{\xi}} \;,
\end{equation}
 with $N_c=3$ denoting the number of colors. This bound automatically
 eliminates large values of $\xi$.
In our analysis we allow for finetuning, hence small values of $\xi$,
and we will not apply this bound. \s

\begin{figure}[t]
\begin{center}
 \hspace*{-0.5cm}\includegraphics[width=8.3cm]{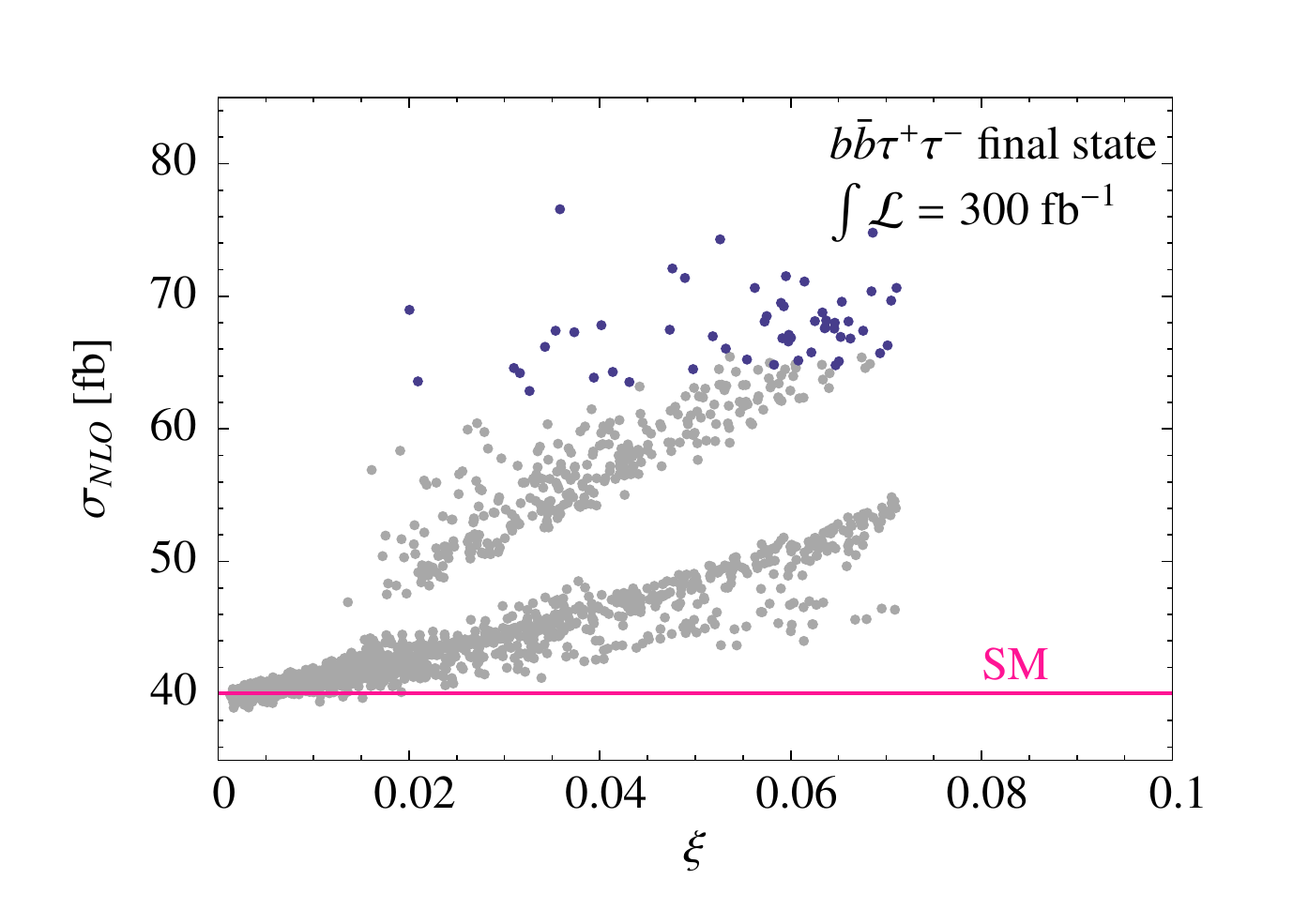}
 \hspace*{0.0cm}
 \includegraphics[width=8.3cm]{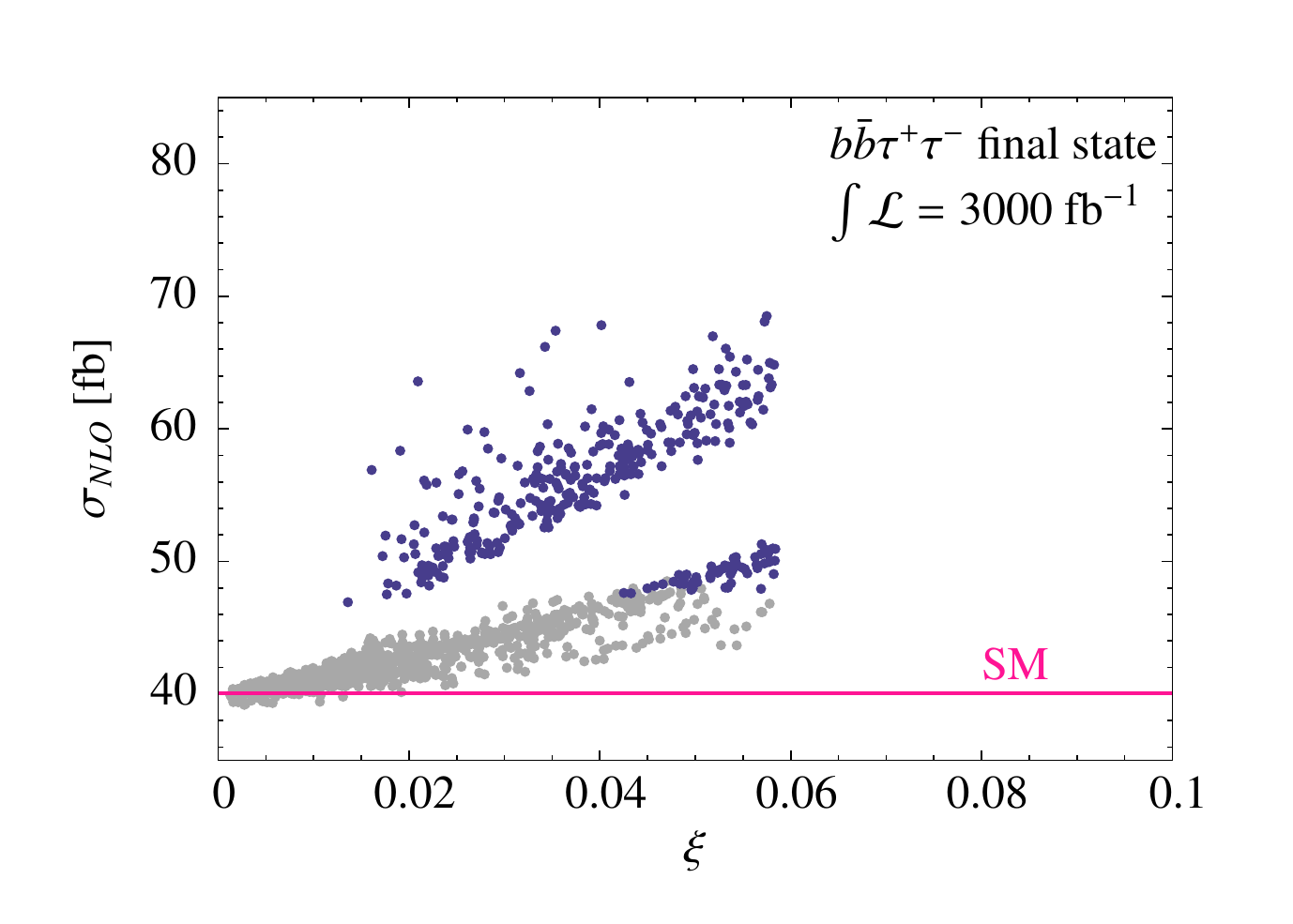}\hspace*{-0.5cm}\\ 
\vspace*{-0.5cm}
  \hspace*{-1cm} \includegraphics[width=8.3cm]{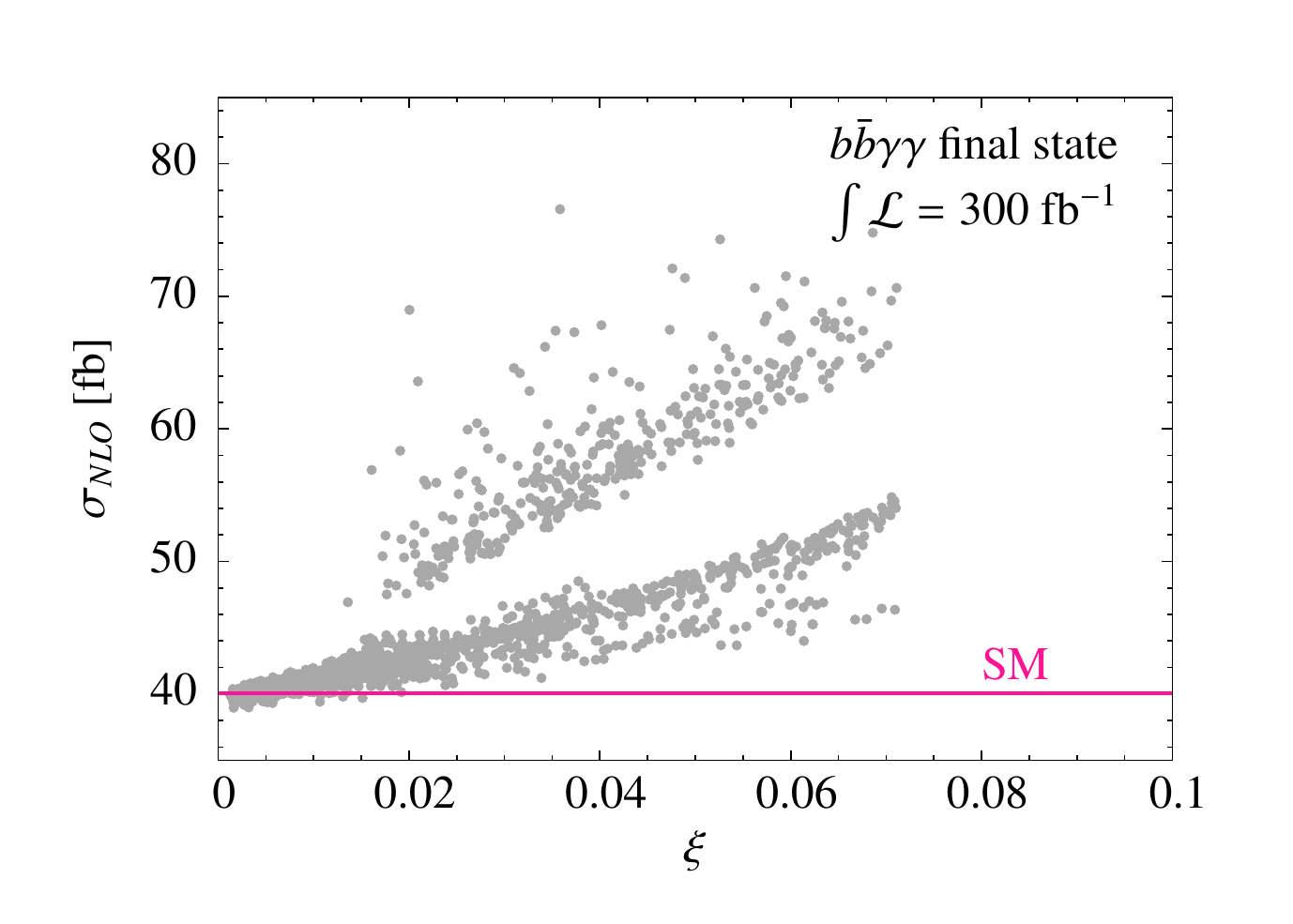}
 \hspace*{0.0cm}
 \includegraphics[width=8.3cm]{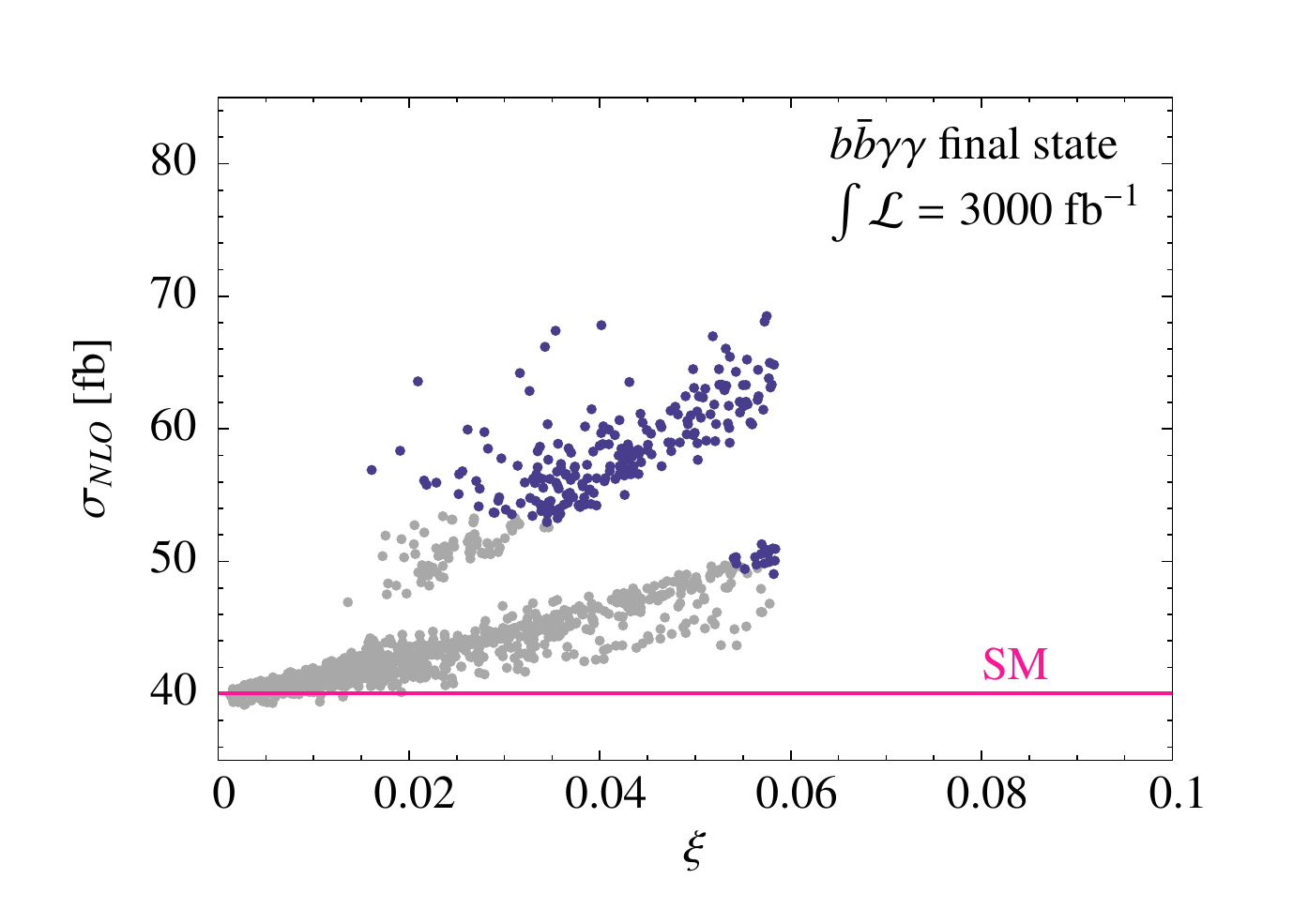}\hspace*{-0.5cm}
 \end{center}
 \vspace*{-0.5cm}
 \caption{The NLO gluon fusion cross section into Higgs pairs in
   MCHM10 for a scan over $\xi$, $y$, $\sin\phi_L$ and $M_{10}$. All
   points pass the EWPTs and, for the respective 
   luminosity, fulfill the projected direct search limits for new fermionic
   resonances at the LHC and allow only for deviations in the Higgs boson
   couplings that are smaller than the according expected sensitivity at the
   LHC. The blue 
   points indicate that the composite Higgs pair production cross
   section can be distinguished from the SM one at 3$\sigma$ in the 
$b\bar{b}\tau^+ \tau^-$ final state (upper) and $b\bar{b}\gamma\gamma$ 
final state (lower) at an integrated luminosity $L=300\text{ fb}^{-1}$
(left)  and $L=3000\text{ fb}^{-1}$ (right). The grey points cannot 
be distinguished from the SM at 3$\sigma$. The pink line is the SM
prediction for the gluon fusion cross section at NLO. \label{fig:MCHM10}}
\end{figure}
For the analysis we performed a scan over the parameter space of the
model by varying the parameters in the range\footnote{Here $\phi_L$ is the
  rotation angle applied on the $SU(2)$ bi-doublet for the 
diagonalization of the mass matrices at LO in $v/f$. It is related to
the parameters of the model by $\tan\phi_L = \lambda_q/(M_{10}+fy/2)$.}
\begin{equation}
 0\leq \xi \leq 1\,, \hspace*{0.5cm} 0< \sin\phi_L\leq 1\,, \hspace*{0.5cm} 
|y|< 4\pi\,, \hspace*{0.5cm} 0\leq M_{10}\leq 10\text{ TeV}\,.
\end{equation}
We excluded points that do not fulfill $|V_{tb}| >0.92$
  \cite{Chatrchyan:2012ep} and the electroweak precision  
tests (EWPTs) at 99\% C.L. using the results of Ref.~\cite{Gillioz:2013pba}.\s

In Fig.~\ref{fig:MCHM10} we show the NLO Higgs pair production cross section via 
gluon fusion as a function of $\xi$. The color code in the plots
indicates whether the points are distinguishable from the SM according
to the criteria given in Eq.~\eqref{crit}, with the blue points being
distinguishable and the grey points not. The upper plots are for the
$b\bar{b}\tau^+ \tau^-$ final state, the lower plots for the
$b\bar{b}\gamma\gamma$ final state, for $\int \mathcal
L=300\text{ fb}^{-1}$ (left) and $\int \mathcal L=3000\text{ fb}^{-1}$
(right), respectively. The upper branch in the plots corresponds to the parameters
$y<0$ and $0<R<1$ with $R=(M_{10}+ f y/2)/M_{10}$. This means that at
LO of the mass matrix expansion in $v/f$, the lightest fermion
partner originates from the $SU(2)$ bi-doublet. The lower branch
corresponds to the cases $y <0$ and $R < 0$ as well as $y> 0$ implying
$R>1$. \s

\begin{figure}[b!]
\begin{center}
 \hspace*{-0.5cm}\includegraphics[width=8.3cm]{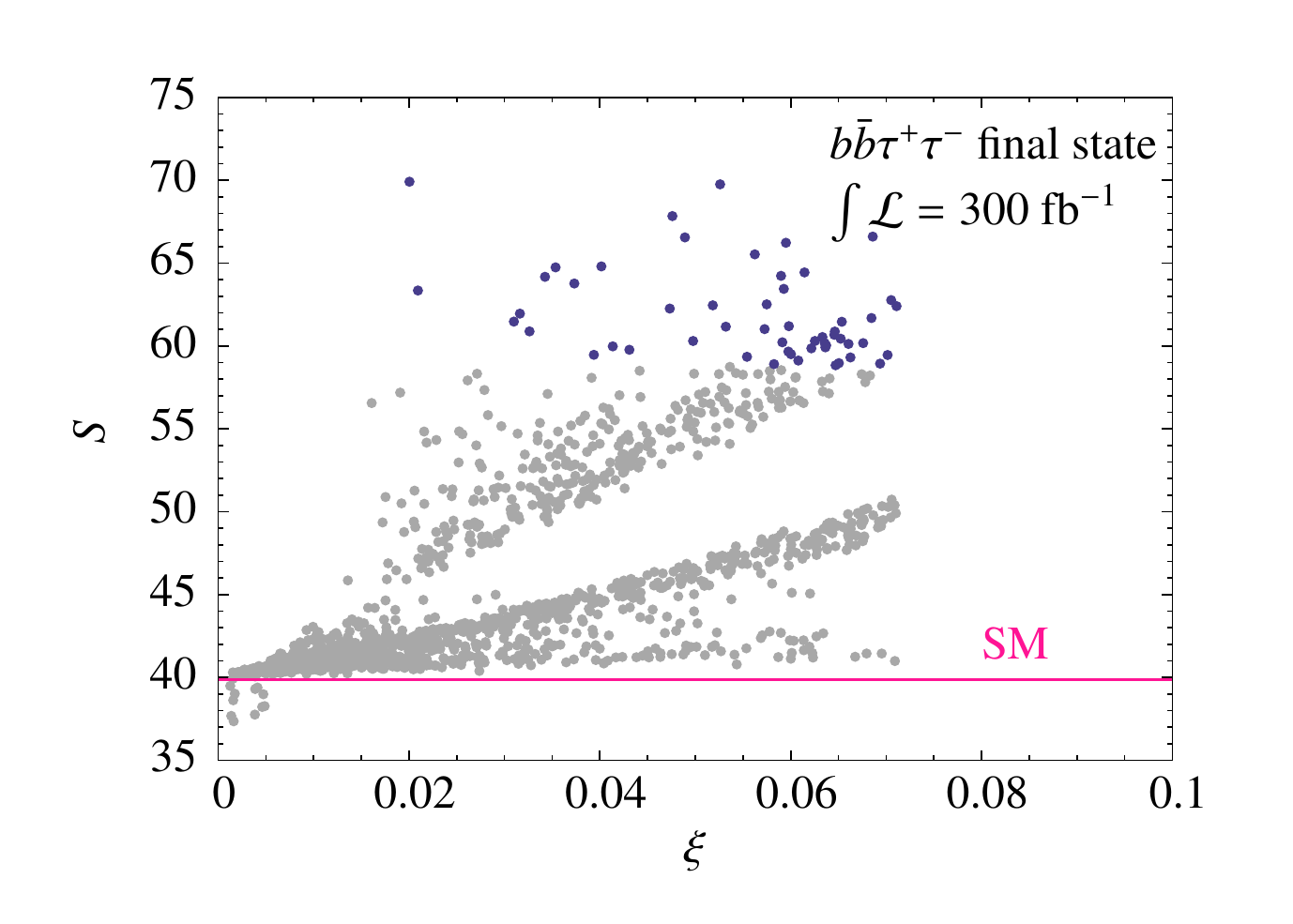}
 \hspace*{0.0cm}
 \includegraphics[width=8.3cm]{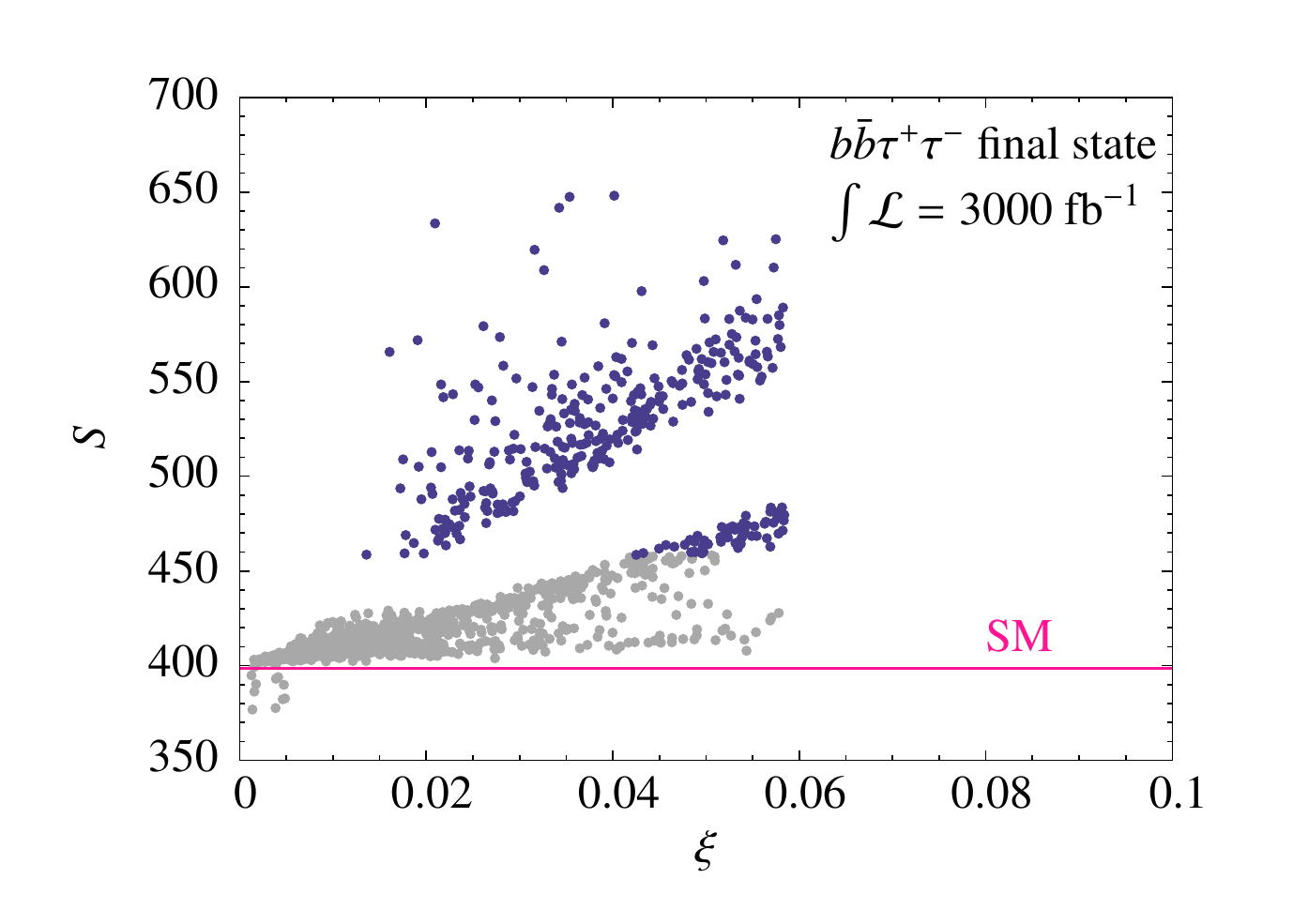}\hspace*{-0.5cm}\\ 
\vspace*{-0.5cm}
  \hspace*{-1cm} \includegraphics[width=8.3cm]{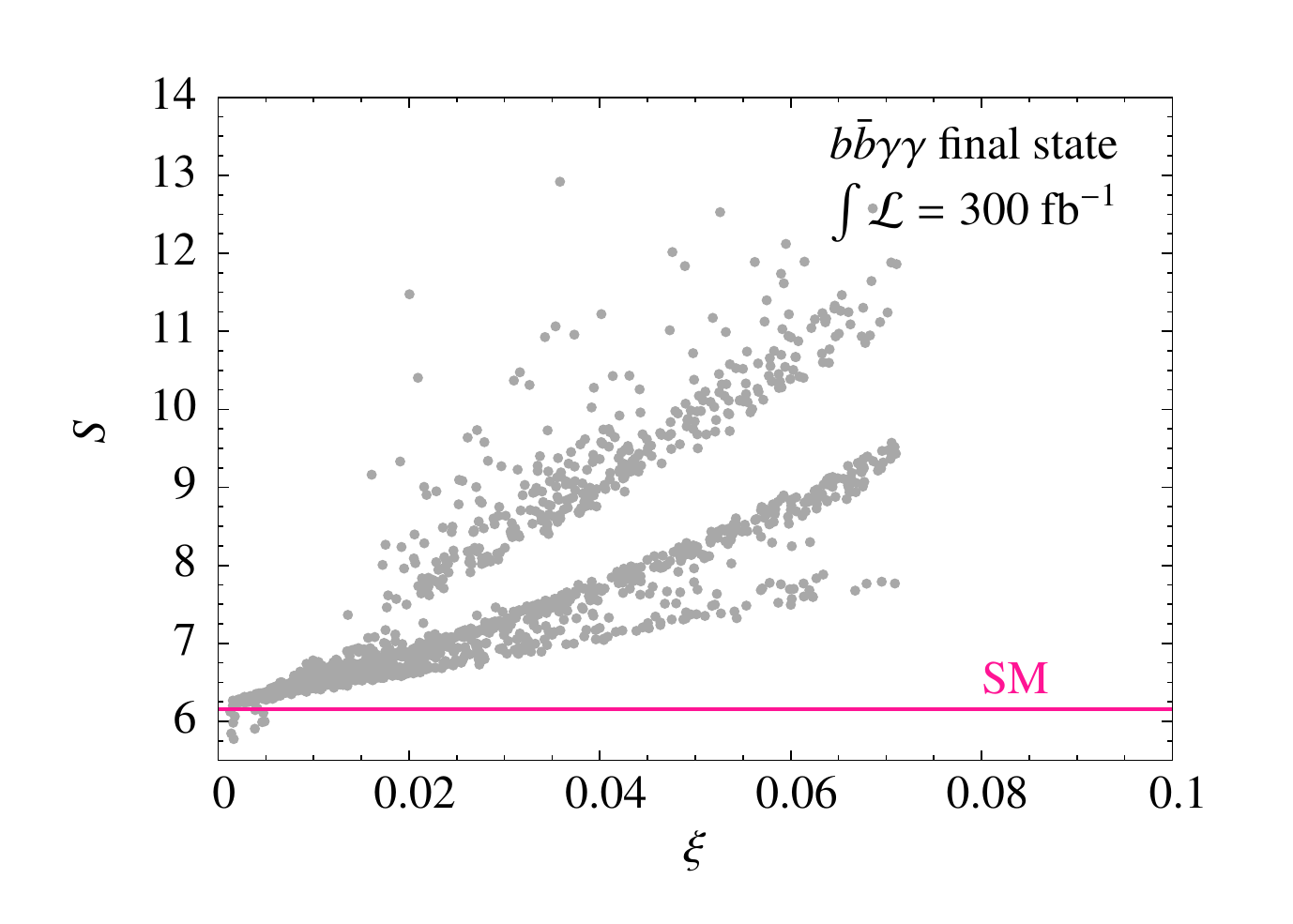}
 \hspace*{0.0cm}
 \includegraphics[width=8.3cm]{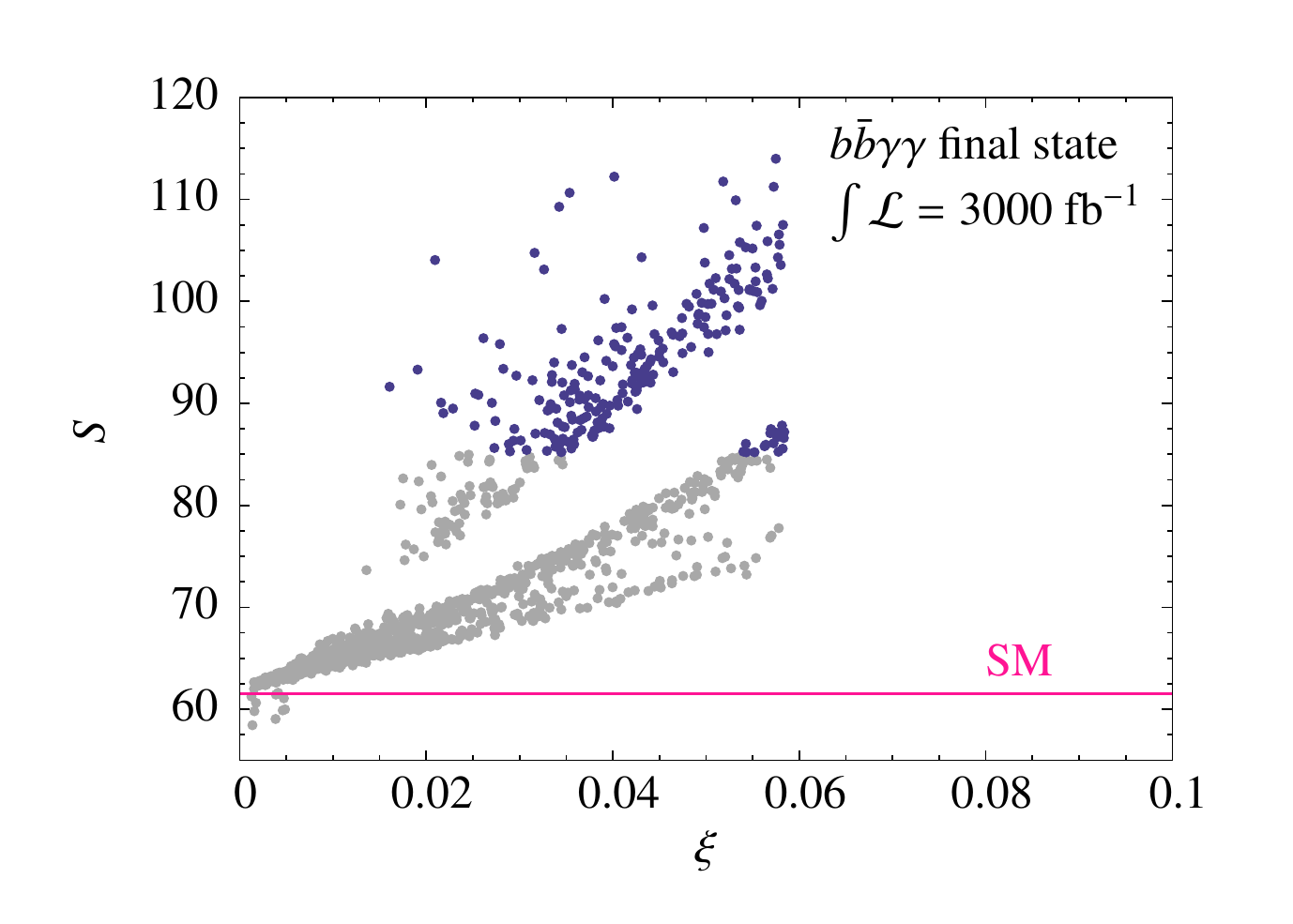}\hspace*{-0.5cm}
 \end{center}
 \vspace*{-0.5cm}
 \caption{Number of signal events $S= \sigma_{NLO} \cdot 2
   \cdot \mbox{BR} (h\to X) 
   \cdot \mbox{BR} (h\to Y) \cdot A \cdot L$ for NLO Higgs pair
   production in the final states $X = b \bar{b}$, $Y=
   \tau^+ \tau^-$ (upper) and $X = b\bar{b}$, $Y= \gamma \gamma$
   (lower) for an integrated luminosity of $L=300$~fb$^{-1}$ (left)
   and $L=3000$~fb$^{-1}$ (right), after multiplication with the individual
   acceptance $A$ of the applied cuts, for the parameter points
   given in Fig.~\ref{fig:MCHM10}. The colour code is the same and the
   pink line refers to the NLO signal rate in the SM. \label{fig:signalnumbers}}
\end{figure}
The plots only show points for which we cannot see new 
physics anywhere else meaning we require that their deviations in the Higgs 
couplings that can be tested at the LHC are smaller than the expected 
sensitivities and that the masses of the new fermionic resonances are above the 
estimated reach of direct searches.\footnote{Note that we did not
    consider possible future results from flavour physics
    and/or the measurement of $V_{tb}$ that could further restrict
    composite Higgs models.} The requirement for only small 
deviations in the Higgs couplings directly restricts the possible values of 
$\xi$ to be smaller than 0.071 and 0.059 for $\int \mathcal L=300\text{ fb}^{-1}$ 
and $\int \mathcal L=3000\text{ fb}^{-1}$,
respectively. 
The value of $\xi$ is restricted more strongly than in
  the MCHM4 due to the different coupling modifications,
  which, considering pure non-linearities, are for the
  Higgs-fermion couplings  $(1-2\xi)/\sqrt{1-\xi}$ in MCHM10 and
  $\sqrt{1-\xi}$ in the MCHM4. Although the interplay of the various
  additional parameters in MCHM10 allows for some tuning in the
  Higgs-bottom (and also Higgs-top) coupling, this is not the case for the
  Higgs-tau coupling. Comparing the MCHM10 with the MCHM5, the Higgs-fermion
  couplings are modified in the same way, barring the effects from the
  additional fermions. The increased number of parameters due to the
  heavy fermions, however, allows for more freedom to accommodate the
  data, so that here the constraint is weaker in the MCHM10. 
The plots show that at $\int \mathcal L=300\text{
  fb}^{-1}$  we cannot expect to discover NP for the first
time in Higgs pair production in the $b\bar{b}\gamma\gamma$
final state, while in the $b\bar{b}\tau^+ \tau^-$ final state NP could
show up for the first time in Higgs pair production. For $\int 
\mathcal L=3000\text{ fb}^{-1}$, we could find both in the 
$b\bar{b}\tau^+ \tau^-$ and the $b\bar{b}\gamma\gamma$ final state points 
which lead to large enough deviations from the SM case to be sensitive
to NP for the first time in Higgs pair production. These
results can be explained with the increased signal rate in the
cases that are sensitive, as can be inferred from
Fig.~\ref{fig:signalnumbers}. The plots show for the parameter points
displayed in Fig.~\ref{fig:MCHM10} the corresponding number of signal 
events for Higgs pair production in the $b\bar{b}\tau^+ \tau^-$ and
$b\bar{b}\gamma \gamma$ final states, respectively, after applying the
acceptance of the cuts
and multiplication with the two
options of integrated luminosity. The blue points clearly deviate by
more than 3$\sigma$ from the SM curve. 

\section{Invariant Mass Distributions \label{sec:invmasssdist}}
Finally, in this section we discuss NP effects in invariant
Higgs mass distributions. The measurement of distributions can give information 
on anomalous couplings \cite{Chen:2014xra} or the underlying
ultraviolet source of NP \cite{Dawson:2015oha}. Even though they are
difficult to be measured due to the small numbers of signal events, they
are important observables for the NP search. In the following we will
show the impact of composite Higgs models on the distributions. \s

\begin{figure}[hb!]
\hspace*{-0.55cm} \includegraphics[scale=0.65]{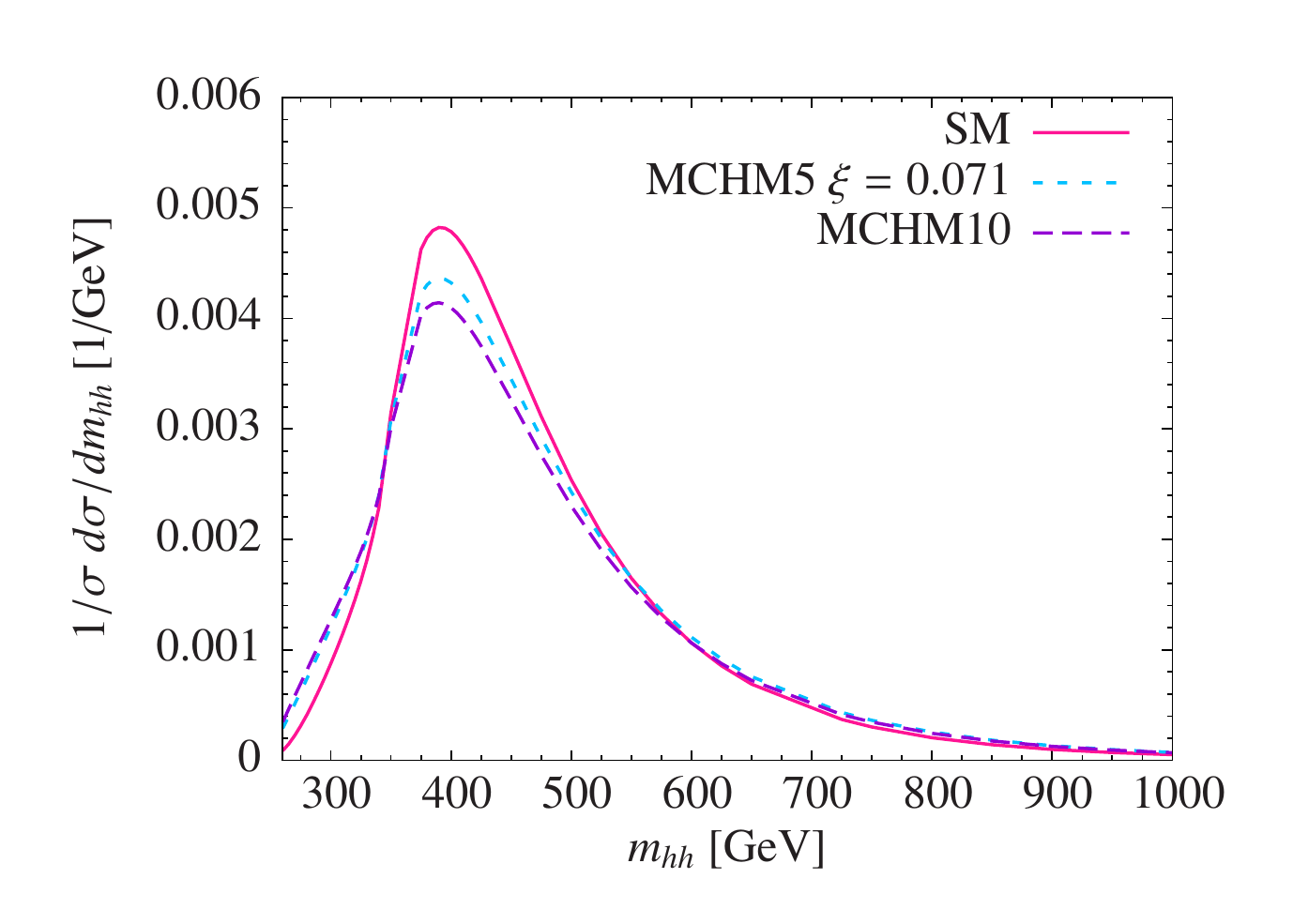}
\hspace*{-0.9cm} 
  \includegraphics[scale=0.65]{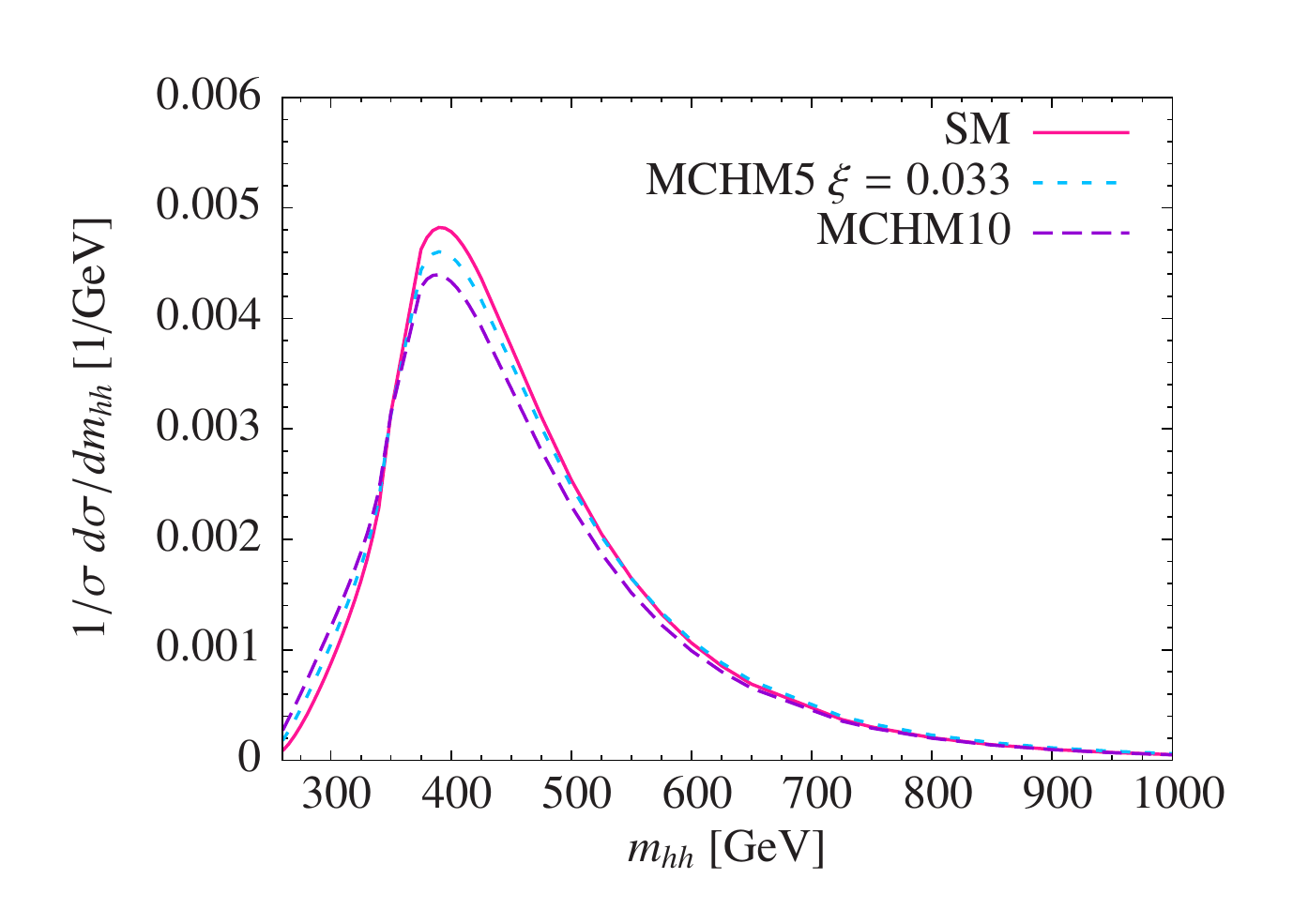}
\vspace*{0cm}
\caption{Normalized invariant mass distributions at NLO QCD for
    the SM (pink solid) and left: MCHM5 with $\xi=0.071$ (blue dotted) and 
    the MCHM10 with fermionic resonances for the parameter choice $\xi=0.071$, 
$y=-7.292$, $\sin\phi_L=0.551$ and $R=0.615$ (violet dashed); right:
MCHM5 with $\xi=0.033$ (blued dotted) and the MCHM10 parameter choice
$\xi=0.033$, $y=-10.898$, $\sin\phi_L= 0.895$ and $R=0.181$ (violet dashed).
\label{fig:dist}}
\end{figure}
Figure~\ref{fig:dist} (left) shows the normalized invariant mass distributions
for the MCHM5 with $\xi=0.071$ (blue dotted) and the MCHM10 with one
multiplet of fermionic resonances for the parameter
 choice $\xi=0.071$, $y=-7.292$, $\sin\phi_L=0.551$ and $R=0.615$
 (violet dashed) compared to the SM case  (pink solid). In the right
 plot we show the same quantity, but for the MCHM5 with $\xi=0.033$
 (blue dotted) and the MCHM10 parameters $\xi=0.033$, $y=-10.898$,
 $\sin\phi_L= 0.895$ and $R=0.181$ (violet dashed) again compared to
 the SM (pink solid).
The cross sections are computed at NLO. Note, however, that the shape of
the invariant mass distributions hardly changes from LO to NLO, since in the LET
approximation the LO cross section mainly factorizes from the
NLO contributions, as discussed in section \ref{sec:NLOQCD}. 
The parameters have been chosen such that in the left plot we allow for a
larger value of $\xi$, while the mass of the lightest top partner of
$m_{T, \text{lightest}}=5441 \text{ GeV}$ is much larger than compared to the case
shown in the right plot with $m_{T, \text{lightest}}=1636 \text{ GeV}$. As
can be inferred from these plots, the largest effect on the
distributions originates from the pure non-linearities, {\it i.e.}~the
value of $\xi$, while the influence of the fermionic resonances on the
shape of the invariant mass distributions is small. Note also that the 
main effect on the distributions emerges from the new $t\bar{t}hh$
coupling. \s

\section{Conclusions \label{sec:concl}}
We presented the NLO QCD corrections to Higgs pair production via gluon 
fusion in the large quark mass approximation in composite Higgs
models with and 
without new heavy fermionic resonances below the UV cut-off. We found that the 
$K$-factor of $\sim 1.7$ is basically independent of the value of
$\xi$ and of the details of the fermion spectrum, as the LO cross
  section dominantly factorizes. The $K$-factor can hence directly be taken over
from the SM to a good approximation. The size of the absolute value of
the cross section, however, sensitively depends on the Higgs
non-linearities and on the fermion spectrum. \s

With the results of our NLO calculation, we 
furthermore addressed the question of whether NP could emerge for the first 
time in Higgs pair production, taking into account the constraints on
the Higgs couplings to SM particles and from direct searches for new
heavy fermions. We focused on composite Higgs
models and found that in simple models where only the Higgs
non-linearities are considered, we cannot expect to be
sensitive to NP for the first time in Higgs pair production. In models with a
multiplet of new fermions below the cut-off, it turned out that there
are regions where NP could indeed be seen for the first time in Higgs
pair production. The subsequent study of the NLO invariant mass
distributions demonstrated, that while there is some sensitivity to
the Higgs non-linearities mainly due to the new 2-Higgs-2-fermion
coupling, the effect of the heavy fermions on the shape of 
the distributions is much weaker. By applying optimized cuts the
sensitivity to new physics effects may possibly be increased in future.

\subsubsection*{Acknowledgments} 
This work is supported in part by the Research Executive Agency (REA) of 
the European Union under Grant No. PITN-GA-2012-316704 (Higgstools).

\section*{Appendix}
\begin{appendix}
\section{Masses and Couplings \label{app:coupmass}}
In the following we give the mass and coupling matrices for the
composite Higgs model MCHM10 with heavy top and fermion partners, needed in
the gluon fusion process into Higgs pairs. With the abbreviations  
\beq 
s_H\equiv \sin(H/f) \;, \qquad c_H \equiv \cos(H/f)
\eeq
and
\begin{equation}
\tilde{m}_a \equiv\frac{1}{4} f y s_H^2+M_{10}\;,\hspace*{1cm}
\tilde{m}_b \equiv \frac{1}{2} f y(1-\frac{1}{2}s_H^2)+M_{10} \;,\hspace*{1cm}
\tilde{m}_c \equiv \frac{1}{2} f yc_H^2+M_{10}\;,
\end{equation}
the terms bilinear in the quark fields of Eq.~\eqref{lagrangian} read 
\small\begin{equation}
-\mathcal{L}_{m_{t}}=
\overline{\left(\begin{array}{c}
t_L\\ u_L\\u_{1L}\\t_{4L}\\T_{4L}
\end{array}\right)}
\left(
\begin{array}{ccccc}
 0 & 0 & 0 & 0 & \lambda_q \\
 0 & \tilde{m}_a& -\frac{1}{4} f
   y s_H^2 & -\frac{1}{4} f y
  c_H s_H & -\frac{1}{4} f y c_Hs_H
   \\
 \lambda_t & -\frac{1}{4} f y s_H^2 & \tilde{m}_a& \frac{1}{4}
f
   yc_H  s_H & \frac{1}{4} f y c_Hs_H
   \\
 0 & -\frac{1}{4} f yc_H s_H & \frac{1}{4} f y c_Hs_H
   & \tilde{m}_b& -\frac{1}{4} f
   y s_H^2 \\
 0 & -\frac{1}{4} f yc_H s_H & \frac{1}{4} f y c_Hs_H
   & -\frac{1}{4} f ys_H^2 &
   \tilde{m}_b
\end{array}
\right) 
\left(\begin{array}{c}
t_R\\ u_R\\u_{1R}\\t_{4R}\\T_{4R}
\end{array} \right)+h.c.\;,\label{eq:app1}
\end{equation}

\vspace*{0.1cm}
\begin{equation}
-\mathcal{L}_{m_{b}}=\overline{ \left(\begin{array}{c}
                      b_L \\d_L \\ d_{1L}\\ d_{4L} 
                      \end{array}\right)}
\left(
\begin{array}{cccc}
 0 & 0 & 0 & \lambda_q \\
 0 &\tilde{m}_a& -\frac{1}{4} f
   ys_H^2 & f y \frac{c_H s_H}{2 \sqrt{2}} \\
 \lambda_b& -\frac{1}{4} f
   ys_H^2  &\tilde{m}_a& -f y
  \frac{c_H s_H}{2 \sqrt{2}} \\
 0 & f y\frac{c_H s_H}{2 \sqrt{2}} & -f y\frac{
  c_H s_H}{2 \sqrt{2}} & \tilde{m}_c
\end{array}
\right)
\left(\begin{array}{c}
                      b_R \\d_R \\ d_{1R}\\ d_{4R} 
                      \end{array}\right)+h.c.\;,\label{eq:app2}
\end{equation}\normalsize

\vspace*{0.1cm}
\noindent
and
\small\begin{equation}
-\mathcal{L}_{m_{\chi}}= \overline{\left(\begin{array}{c}
                          \chi_L\\\chi_{1L}\\\chi_{4L}
                         \end{array}\right)}
\left(
\begin{array}{ccc}
\tilde{m}_a& -\frac{1}{4} f
   ys_H^2 & f y\frac{ c_H s_H}{2 \sqrt{2}} \\
 -\frac{1}{4} f ys_H^2 &
  \tilde{m}_a & -f y\frac{
  c_H s_H}{2 \sqrt{2}} \\
 f y\frac{c_H s_H}{2 \sqrt{2}} & -f y\frac{
  c_H s_H}{2 \sqrt{2}} & \tilde{m}_c
\end{array}
\right)
\left(\begin{array}{c}
                          \chi_R\\\chi_{1R}\\\chi_{4R}
                         \end{array}\right)+h.c.\;.\label{eq:app3}
\end{equation}\normalsize
Shifting the Higgs field in Eqs.~\eqref{eq:app1}--\eqref{eq:app3},
encoded in $s_H$ and $c_H$, respectively, by its VEV, 
{\it i.e.}~$H = \langle H\rangle + h$, leads to the mass matrices
$M_{t}$, $M_{b}$ and $M_\chi$. They can be diagonalized by a bi-unitary
transformation
\begin{equation}
 \left(U_L^{(t/b/\chi)}\right)^{\dagger}M_{(t/b/\chi)}U_R^{
(t/b/\chi) } =M_ {(t/b/\chi)}^{\text{diag}}\;,\label{eq:app4}
\end{equation}
where $U_{L,R}^{(t/b/\chi)}$ denote the transformations that
diagonalize the mass matrix in the top, bottom and charge-5/3 ($\chi$)
sector, respectively. Expansion of the mass matrices
Eqs.~\eqref{eq:app1}--\eqref{eq:app3} in the interaction eigenstates up to
first order in the Higgs field leads to the Higgs coupling matrices
$\tilde{G}_{ht \bar{t}}$ to a pair of charge-2/3 quarks and
$\tilde{G}_{hb \bar{b}}$ to a quark pair of charge
$-1/3$, respectively, given by
\small\begin{equation} 
-\mathcal{L}_{h t\bar{t}}=
y\;h\overline{\left(\begin{array}{c}
t_L\\ u_L\\u_{1L}\\t_{4L}\\T_{4L}
\end{array}\right)}
\underbrace{\left(\begin{array}{ccccc}
 0 & 0 & 0 & 0 & 0 \\
 0 & \frac{1}{2}s_H c_H& - \frac{1}{2}s_H
c_H
   & \frac{1}{4}(2 s_H^2-1) 
 &  \frac{1}{4}(2 s_H^2-1) 
   \\
 0& - \frac{1}{2} s_H c_H&
\frac{1}{2}s_H c_H&\frac{1}{4}(1-2 s_H^2)  &
\frac{1}{4}(1-2 s_H^2)
   \\
 0 & \frac{1}{4}(2 s_H^2-1)& \frac{1}{4}(1-2 s_H^2)
   & -\frac{1}{2}s_H c_H&
-\frac{1}{2}s_H c_H\\
 0 & \frac{1}{4}(2 s_H^2-1) & \frac{1}{4}(1-2 s_H^2)
   & - \frac{1}{2}s_H c_H&
   -\frac{1}{2}s_H c_H
\end{array}\right)}_{\tilde{G}_{ht\bar{t}}/y}
\left(\begin{array}{c}
t_R\\ u_R\\u_{1R}\\t_{4R}\\T_{4R}
\end{array} \right)_{H=\langle H\rangle}\hspace*{-1cm} +h.c.\;,\label{mod11}
\end{equation}
\begin{equation}
-\mathcal{L}_{hb\bar{b}}=y\;h\overline{ \left(\begin{array}{c}
                      b_L \\d_L \\ d_{1L}\\ d_{4L} 
                      \end{array}\right)}
\underbrace{\left(
\begin{array}{cccc}
 0 & 0 & 0 & 0 \\
 0 &\frac{1}{2}s_H c_H& -\frac{1}{2}
   s_H c_H & \frac{1}{2 \sqrt{2}}(1-2 s_H^2) \\
 0&-\frac{1}{2}
   s_H c_H&\frac{1}{2}s_H c_H& 
  \frac{1}{2 \sqrt{2}}(2 s_H^2-1) \\
 0 & \frac{1}{2 \sqrt{2}}(1-2 s_H^2) & \frac{
  1}{2 \sqrt{2}}(2 s_H^2-1) & -s_H c_H
\end{array}
\right)}_{\tilde{G}_{h b\bar{b}}/y}
\left(\begin{array}{c}
                      b_R \\d_R \\ d_{1R}\\ d_{4R} 
                      \end{array}\right)_{H=\langle
H\rangle}\hspace*{-1cm}+h.c.\;.\label{mod12}
\end{equation}\normalsize
Expansion up to second order in the Higgs field yields the
2-Higgs-2-fermion coupling matrices $\tilde{G}_{hht\bar{t}}$ and 
$\tilde{G}_{hh b\bar{b}}$, that can be cast into the form
\small\begin{equation} 
-\mathcal{L}_{h h t\bar{t}}=
\frac{y}{2\,f}\;h\overline{\left(\begin{array}{c}
t_L\\ u_L\\u_{1L}\\t_{4L}\\T_{4L}
\end{array}\right)}
\underbrace{\left(\begin{array}{ccccc}
 0 & 0 & 0 & 0 & 0 \\
 0 &(1-2 s_H^2)& - (1- 2 s_H^2)
   & 2 s_H c_H
 &  2 s_H c_H
   \\
 0& - (1- 2 s_H^2) &
(1- 2 s_H^2)&- 2 s_H c_H&
\frac{1}{4}(1-2 s_H^2)
   \\
 0 & 2 s_H c_H& -2 s_H c_H
   & -(1- 2 s_H^2)&
-(1- 2 s_H c_H)\\
 0 & 2 s_H c_H& 2 s_H c_H
   & - (1- 2 s_H^2)&
   -(1- 2 s_H^2)
\end{array}\right)}_{2\, f\, \tilde{G}_{hht\bar{t}}/y}
\left(\begin{array}{c}
t_R\\ u_R\\u_{1R}\\t_{4R}\\T_{4R}
\end{array} \right)_{H=\langle H\rangle}\hspace*{-1cm} +h.c.\;,\label{mod11a}
\end{equation}
\begin{equation}
-\mathcal{L}_{hb\bar{b}}=\frac{y}{2 f}\;h\overline{ \left(\begin{array}{c}
                      b_L \\d_L \\ d_{1L}\\ d_{4L} 
                      \end{array}\right)}
\underbrace{\left(
\begin{array}{cccc}
 0 & 0 & 0 & 0 \\
 0 &(1- 2 s_H^2)& -(1- 2 s_H^2) & -\sqrt{2}s_H c_H\\
 0&-(1- 2 s_H^2)& (1- 2 s_H^2)& 
  \sqrt{2}s_H c_H \\
 0 & -\sqrt{2}s_H c_H & \sqrt{2} s_H c_H & -2 (1- 2 s_H^2)
\end{array}
\right)}_{2\, f\, \tilde{G}_{h h b\bar{b}}/y}
\left(\begin{array}{c}
                      b_R \\d_R \\ d_{1R}\\ d_{4R} 
                      \end{array}\right)_{H=\langle
H\rangle}\hspace*{-1cm}+h.c.\;.\label{mod13}
\end{equation}\normalsize
The coupling matrices in the mass eigenstate basis are obtained by
rotation with the unitary matrices defined in Eq.~\eqref{eq:app4}, {\it
  i.e.}~($q=t,b$)
\beq
(U_L^q)^\dagger \tilde{G}_{hq\bar{q}} U_R^q = G_{hq\bar{q}} \qquad
\mbox{and} \qquad
(U_L^q)^\dagger \tilde{G}_{hhq\bar{q}} U_R^q = G_{hhq\bar{q}} \;.
\label{eq:tbcoups}
\eeq

\end{appendix}


\end{document}